\documentclass[11pt]{article}
\usepackage{color}
\usepackage{graphicx}
\usepackage{amssymb}
\usepackage{wasysym}

\newsavebox{\PSLASH}
\sbox{\PSLASH}{$p$\hspace{-1.8mm}/}

\newcommand{\arcsinh}{\mathrm{arcsinh}}
\newcommand{\arctanh}{\mathrm{arctanh}}
\newcommand{\de}{\mathrm{d}\hspace{0.08em}}

\begin{document}
\title{Ashkin-Teller model on the iso-radial graphs}
\author{S.~Lottini$^1$\footnote{e-mail: s.lottini@uni-muenster.de},
	M.~A.~Rajabpour$^2$\footnote{e-mail: rajabpour@to.infn.it} \\ \\
$^1$ Westf\"alische Wilhelms-Universit\"at M\"unster, \\
Wilhelm-Klemm-Stra\ss{}e 9, 48149 M\"unster, Germany \\
$^2$ Dip.~di Fisica Teorica \& INFN, Universit\`a di
Torino, \\ Via P.~Giuria 1, 10125 Torino, Italy }
\maketitle
\begin{abstract}
We find the critical surface of the Ashkin-Teller model on the generic iso-radial graphs by using the results for the
anisotropic Ashkin-teller model on the square lattice.
Different geometrical aspects of this critical surface are discussed, especially their connection to the anisotropy angle.
The free energy of the model on the generic iso-radial graph is extracted using the inversion identities.
In addition, lattice holomorphic variables are discussed at some particular points of the critical line.
We check our conjectures numerically for the anisotropic triangular-lattice Ashkin-Teller model.
\end{abstract}
\vspace{5mm}
\begin{small}
\end{small}

\section{\label{sec:introduction}Introduction}
The Ashkin-Teller model, as a simple spin model exhibiting a critical line in its phase diagram, has been studied for many years.
This model was introduced by Ashkin and Teller \cite{Ashkin}, and can be seen as two coupled Ising models.
Some physical realisations of the Ashkin-Teller model are layers of  molecules
adsorbed on surfaces, such as selenium on nichel, molecular oxygen on graphite, and atomic oxygen on tungsten \cite{DR,bak, BEW}. The model can be
mapped, on the square lattice, to the staggered $6$-vertex
model, and the phase diagram and
some of the critical exponents were found by mapping the
model to the Solid-on-Solid (SOS) model \cite{wu_and_lin, Kadanoff_and_Brown,baxter,nienhuis2}. Although the phase diagram is qualitatively quite well known, the model has not been solved exactly for the generic coupling constants as the staggered $6$-vertex model has not been solved so far. 
The staggering of the model disappears on the critical line, where one can completely solve the model \cite{baxter}. The quantum version of the Ashkin-Teller model was also studied by considering the highly anisotropic model on the square lattice and has a different phase diagram \cite{Kohmoto_den_nijs}. The continuum version of the model is as much complicated as the discrete version. It is related to
the sine-Gordon model \cite{Kohmoto_den_nijs} on the critical line, see \cite{DG} for more recent developments.
On the critical line one can also write the action with respect to fermionic variables (it is the massless Thirring model \cite{Den_nijs}).
The corresponding conformal field theory of the model on the critical line, called orbifold conformal field theory, 
was discussed in \cite{yang,Ginsparg,Saleur,Schellekens,DVVV}, see also \cite{von_gehlen_and_rittenberg} . The four-point correlation functions of the spin operators were also calculated exactly 
by conformal field theory methods \cite{zamolodchikov}. Although in the physics community it is widely accepted that the above continuum models describe
the critical line of the Ashkin-Teller model, there is no rigorous proof for the validity of the results.

There was no systematic way to 
study the continuum limit, even for simpler systems such as percolation or Ising models, before the invention of Schramm-Loewner evolution by Schramm \cite{schramm}, for review see \cite{Kager}.
This opened the way to investigate the continuum versions of the critical lines in statistical models. Recently Chelkak and Smirnov
showed that this method is also useful to investigate the universality of statistical models. They proved that the Ising model on different iso-radial graphs, which will be discussed in Section \ref{sec:Ising_Model_on_the_Iso-radial_Graphs}, have the same continuum limit at criticality \cite{CS}. The iso-radial graphs 
are important because  they are the most generic graphs preserving some properties of complex analysis and have a well-defined continuum limit.
They are also the most general graphs where one can expect conformal invariance, in the thermodynamic limit, for critical statistical models.

These reasons are enough to stimulate theoretical physicists to investigate statistical models on iso-radial graphs. 
The study of statistical models on planar graphs was initiated by Baxter \cite{baxter2,baxter3}. He introduced the most generic planar graphs such
that the star-triangle relations (or Yang-Baxter equations) are well-defined for the eight vertex model. These graphs, now called Baxter lattices,
  are those made of the union of simple non self-intersecting curves crossing the plane, with the property that  a given curve can intersect any of the others
at most once. Iso-radial graphs are a special embedding of the Baxter lattices. Baxter solved the eight-vertex model \cite{baxter2} and 
the Potts model \cite{baxter3} on these graphs. 
With the above motivations about the importance of iso-radial graphs, we will study the critical line of the Ashkin-Teller model on them.
Most of the studies on the Ashkin-Teller model are performed on the square lattice. To the best of our knowledge the formula for the critical line of the Ashkin-Teller model on
the generic iso-radial graph is not written anywhere.

In this paper, by using the formula for the critical line of the anisotropic Ashkin-Teller model
on the square lattice, we will find the critical surface on iso-radial graphs.
The central claim is, the formula 
for the critical point of the model on the anisotropic square lattice is sufficient to find the critical point on the most generic iso-radial graph.
This was already checked for the Ising model in \cite{mercat} and for the Fateev-Zamolodchikov point of $\mathbb{Z}_N$ models \cite{ZF} in \cite{RC}. Here we will investigate 
its validity for the critical line of the Ashkin-Teller model. The paper is organised as follows: in Section \ref{sec:definition} we define the Ashkin-Teller model and recall its different
phases on the square lattice; in Section \ref{sec:Ising_Model_on_the_Iso-radial_Graphs} we first review the general method for the Ising model with arbitrary couplings on iso-radial graphs;
in Section \ref{sec:Anisotropic_Ashkin-Teller}, by employing the 
method of Section \ref{sec:Ising_Model_on_the_Iso-radial_Graphs}, we introduce our conjecture
for the critical line on iso-radial graphs, then, in Section \ref{sec:montecarlo_triangular}, we provide numerical calculations for the anisotropic triangular lattice in support of it. Finally, in the last Section
we will conclude our findings and also discuss possible future investigations.

\section{\label{sec:definition}Ashkin-Teller model: definition} 
\setcounter{equation}{0}
The partition function of the Ashkin-Teller model is given by:
\begin{eqnarray}\label{ATZ1}
    Z  &=&  \sum_{C}\prod_{i,j}W(i,j)=\sum_{C}\prod_{i,j}e^{S_{ij}}\;, \nonumber\\
  S_{ij}&=&\beta(\sigma_{i}\sigma_{j}+\tau_{i}\tau_{j})+\alpha (\sigma_{i}\sigma_{j}\tau_{i}\tau_{j})\;,
\end{eqnarray}
where $\sigma_{i}$ and $\tau_{i}$ are Ising variables, $i$ and $j$ run over all pairs of neighbouring sites, and the summation is over all possible configurations. One can write the above partition function with respect to complex variables as follows:
\begin{equation}\label{ATZ2}
    Z  =  \sum_{C}\prod_{i,j}W(i,j)= \sum_{C}\prod_{i,j}c_{0}\big(1+x_{1}s_{i}^{*}s_{j}+x_{2}s_{i}^{*2}s_{j}^{2}+x_{1}s_{i}^{*3}s_{j}^{3})\;,
\end{equation}
where $s_{i}=\frac{\sigma_{i}+i\tau_{i}}{\sqrt{2}}$ and 
\begin{eqnarray}\label{ATZ3}
c_{0}&=&1+\tanh \alpha \tanh^{2}\beta\;,\\
x_{1}&=&\frac{e^{2\beta}-e^{-2\beta}}{e^{2\beta}+e^{-2\beta}+2e^{-2\alpha}}\;,\\
x_{2}&=&\frac{e^{2\beta}+e^{-2\beta}-2e^{-2\alpha}}{e^{2\beta}+e^{-2\beta}+2e^{-2\alpha}}\;.
\end{eqnarray}
In this notation the $\mathbb{Z}_{4}$ symmetry of the Ashkin-Teller model is evident. Since some of the forthcoming formulas have a simpler form in one or the other of 
the above notations, we will mostly provide them in both notations. The duality transformation in the $\beta$, $\alpha$ notation is implemented by the relations:
\begin{eqnarray}\label{duality1}
\tilde{\beta}(\beta,\alpha)&=&\frac{1}{4}\log\Big(\frac{1+\tanh^{2}\beta \tanh \alpha}{\tanh^{2}\beta+\tanh \alpha}\Big)\;, \nonumber\\
\tilde{\alpha}(\beta,\alpha)&=&\frac{1}{4}\log\Big(\frac{(\coth\beta+\tanh\beta\tanh\alpha)(\coth\beta+\tanh\beta\coth\alpha)}{2+\tanh\alpha+\coth\alpha}\Big)\;.
\end{eqnarray}
In the $x_{1}$, $x_{2}$ notation they become simply:
\begin{eqnarray}\label{duality2}
\widetilde{x}_{1}&=&\frac{1-x_{2}}{1+x_{2}+2x_{1}}\;, \nonumber\\
\widetilde{x}_{2}&=&\frac{1+x_{2}-2x_{1}}{1+x_{2}+2x_{1}}\;.
\end{eqnarray}
On the arbitrary lattice, the physical region of the model (i.~e.~that with positive Boltzmann weights) is the triangle enclosed by the lines
\begin{equation}\label{physical_region}
1+2x_{1}+x_{2}\geq 0\;,\hspace{1cm}1-2x_{1}+x_{2}\geq 0\;, \hspace{1cm}x_{2}\leq1\;.
\end{equation}
To stick to the ferromagnetic region we also need to consider $x_{1}\geq0$. From Eq.~(\ref{duality2}) one can easily find the self-dual line $x^{S}_{2}+2x^{S}_{1}=1$; on the square lattice, this line is critical for $x_{1}\geq \frac{1}{3}$.
On the square lattice one can write the following equations for the correlation exponent \cite{baxter}:
\begin{eqnarray}\label{AT_R}
\frac{1}{\nu}&=&2-\frac{2}{r^{2}}\;,\\
\sin\Big(\frac{\pi r^{2}}{8}\Big)&=&\frac{1}{2}\Big(\frac{1}{x_{1}}-1\Big)\;.
\end{eqnarray}
It is useful to identify some 
well-known points on the critical line. The first interesting point is the 4-state Potts model. 
On the square lattice it is given by the following values in the $\beta$-$\alpha$ and $x_{1}$-$x_{2}$ planes:
\begin{eqnarray}\label{Potts_point_square}
\beta^{S}_{P}&=&\alpha^{S}_{P}=\frac{1}{4}\log 3\simeq0.274653\;,\\
x^{S}_{1P}&=&x^{S}_{2P}=\frac{1}{3}\;,\\
\nu_{P}&=&\frac{2}{3}\;.
\end{eqnarray}
The next interesting point is the Fateev-Zamolodchikov (F-Z) point \cite{ZF}, with the following values:
\begin{eqnarray}\label{FZ_square}
\beta^{S}_{FZ}&=&-\frac{1}{8}\log\Big(17-8\sqrt{2}+(8-2\sqrt{2})(\sqrt{2-\sqrt{2}}-\sqrt{2+\sqrt{2}})\Big)\simeq 0.302923\;,\nonumber\\
\alpha^{S}_{FZ}&=& \frac{1}{4}\log(\sqrt{2}+1)\simeq 0.220343\;,\nonumber\\
x^{S}_{1FZ}&=&\frac{\sin(\frac{\pi}{16})}{\sin(\frac{3\pi}{16})} \simeq 0.351153\;,\\
x^{S}_{2FZ}&=&\frac{\sin(\frac{\pi}{16})}{\sin(\frac{3\pi}{16})}\frac{\sin(\frac{5\pi}{16})}{\sin(\frac{7\pi}{16})}\simeq 0.297693\;,\nonumber\\
\nu_{FZ}&=&\frac{3}{4}\;. \nonumber
\end{eqnarray}
The third interesting point is the Ising point; it is the critical point along the Ising line $\alpha=0$ (or $x_{2}=x^{2}_{1}$), and has the following values:
\begin{equation}
\begin{array}{rclrcl}
\beta^{S}_{I}&=&\frac{1}{2}\log(1+\sqrt{2})\simeq 0.440687\;, & \alpha^{S}_{I} &= & 0\;, \\
x^{S}_{1I}&=&\sqrt{2}-1\;, & x^{S}_{2I} &=& 2-\sqrt{2}\;, \\
\nu_{I}&=&1\;. & & & 
\end{array}
\label{Ising_point_square}
\end{equation}
The next interesting point is the one lying on the boundary of the physical region; we call it the ``terminator'' point:
 \begin{eqnarray}
\beta^{S}_{T} & = & - \alpha^{S}_{T} + \log \sqrt{2} \to \infty \;,\label{Terminator_point_square_b}
\\x^{S}_{1T}&=&\frac{1}{2}\;,\;\;x^{S}_{2T}=0\;,\label{Terminator_point_square_x}\\
\nu_{T}&=&2\;.\label{Terminator_point_square_nu}
\end{eqnarray}
On the square lattice, the $X$-$Y$ point (where $\nu\rightarrow \infty$) is not in the physical region:
if we naively demand this value of the correlation exponent from the equation (\ref{AT_R}) then we will get:
\begin{equation}
x^{S}_{1XY} = \frac{1}{1+2\sin(\frac{\pi}{8})}\simeq0.566454\;,\;\;\;
	x^{S}_{2XY} = \frac{2\sin(\frac{\pi}{8})-1}{2\sin(\frac{\pi}{8})+1}\simeq -0.132909\;, \\
\label{XY_point_square_x1x2}
\end{equation}
but the translation to the $\beta$-$\alpha$ plane poses some problems, as one gets complex-valued couplings:
\begin{equation}
\beta^{S}_{XY} = \frac{1}{4}\log\Big[-\Big(1+2\sqrt{2}+2(1+\sqrt{2})\sqrt{2-\sqrt{2}}\Big)\Big]\;,\;\;
\alpha^S_{XY} = \frac{1}{4}\log\big(-(\sqrt{2}-1)\big)\;.
\label{eq:XY_point_square_ba}
\end{equation}
We will conclude here the introduction of particular points on the square-lattice critical line; in Appendix \ref{app:triang_honeycomb} we will locate those points on the triangular and honeycomb lattices and provide more details.
\begin{figure}
\begin{center}
\includegraphics[angle=0,scale=0.44]{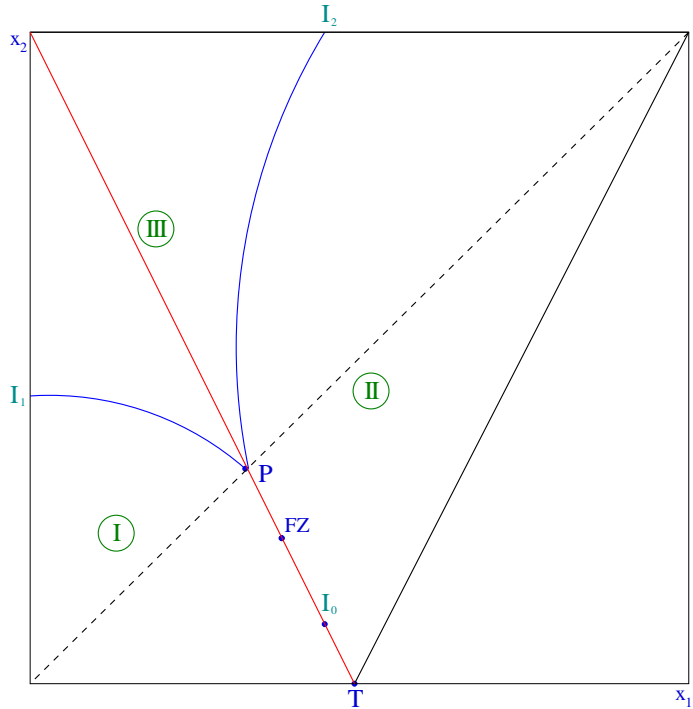}
\caption{The phase diagram of the Ashkin-Teller model. In region I we have $\langle s\rangle\neq 0$ and $\langle s^{2}\rangle\neq 0$, in region II we have $\langle s\rangle=\langle s^{2}\rangle=0$, and finally in region III $\langle s\rangle=0$ and $\langle s^{2}\rangle\neq 0$. The blue lines (mapped onto each other by duality) are Ising lines separating region II from I and III. Thick black lines are the boundary of the physical region and the red line is the self-dual line: it is critical between points P and T. $I_{o}$, $I_{1}$ and $I_{2}$ are Ising points.}
\end{center}
\end{figure}

\section{\label{sec:Ising_Model_on_the_Iso-radial_Graphs}Ising model on iso-radial graphs}
\setcounter{equation}{0}
In this Section we  review the method employed in \cite{mercat,RC} to extract the critical point of the Ising model on iso-radial graphs
 starting from the critical point of the anisotropic square-lattice Ising model. Here we generalize the situation to the off-critical region as well. Then, in the next Section, by using 
the solution for the anisotropic square-lattice Ashkin-Teller model, we will derive the critical surface of this model on the generic iso-radial graph.

To see the importance of the anisotropy angle in the Ising model, consider a square lattice with different couplings in the $x$ and $y$
directions. To fix the notation write the partition function as:
 \begin{equation}\label{Ising_model_partition_function}
Z=\sum_{C}\prod_{x,y,x',y'}W_{I}(x,y,x',y')=\sum_{C}\prod_{x,y,x',y'}(1+w_{x}s_{x,y}s_{x',y}+w_{y}s_{x,y}s_{x,y'})\;,
\end{equation}
where the primed coordinates denote nearest-neighbours of $(x,y).$
In this case, in order to maintain invariance under lattice translations and reflection symmetry 
about the $x$ and $y$ axes, the only allowed transformations are relative scalings of the $x$ and $y$  \cite{BPP}. In our case each elementary 
square of the covering lattice is deformed into a rectangle. To proceed, we define disorder variables $\mu_{\tilde{r}}$ on the sites of the dual lattice, 
which is another rectangular lattice with vertices at the centers $\tilde{r}$ of the faces of the original lattice.
 The insertion of a disorder variable corresponds to modifying the weights so that the order variable $s_{r}$ has monodromy $s_{r}\rightarrow-s_{r}$
on taking the point $r$ in a closed circuit around $\tilde{r}$. This is equivalent to the introduction of a path on the sites of the dual lattice from 
$\tilde{r}$ to infinity, such that the weights on edges $(rr')$ intersected by the path are modified by the substitution $s_{r}s_{r'}\rightarrow -s_{r}s_{r'}$.
To fix the gauge we will consider a path parallel to the $x$ axis and going to $+\infty$.
Thus the disorder operator has the following form for the Ising model:
\begin{equation}\label{Ising_model_disorder_definition}
\mu_{\tilde{r}}=\prod_{_{(rr')\rm{intersected\hspace{1mm} by\hspace{1mm}path}}}\frac{(1-w_{rr'}s_{r}s_{r'})}{(1+w_{rr'}s_{r}s_{r'})}\;.
\end{equation}
It is easy to see that the following equation is valid for two adjacent disorder variables:
\begin{equation}\label{Ising_model_disorder_relation}
\mu_{\tilde{1}}=\frac{1-w_{y}s_{1}s_{2}}{1+w_{y}s_{1}s_{2}}\mu_{\tilde{2}}\;.
\end{equation}
Now consider the quadrilateral whose vertices are the two neighbouring spin variables $s_{1}$ and $s_{2}$ and the two neighbouring disorder variables 
$\mu_{\tilde{1}}$ and $\mu_{\tilde{2}}$: this quadrilateral is living on the  covering lattice. Using Eq.~(\ref{Ising_model_disorder_relation}) one can show that:
\begin{equation}\label{equation_of_motion_1}
-\psi_{1\tilde{1}}+e^{-i(\pi-\theta)}\psi_{2\tilde{1}}
+\psi_{2\tilde{2}}+e^{i\theta}\psi_{1\tilde{2}}=\Big(e^{i\theta}+\frac{w_{y}-i}{w_{y}+i}e^{i\frac{\theta}{2}}\Big)(\psi_{1\tilde{2}}-\psi_{2\tilde{1}})\;,
\end{equation}
with spinor-like variables $\psi$ defined as:
\begin{equation}
\begin{array}{rclrcl}
\psi_{1\tilde{1}}&=&e^{i\frac{\pi}{2}}s_{1}\mu_{\tilde{1}}\;,
	& \psi_{2\tilde{1}} &=& e^{-i\frac{\theta}{2}}s_{2}\mu_{\tilde{1}}\;,\\
\psi_{1\tilde{2}}&=& e^{-i\frac{(\pi-\theta)}{2}}s_{1}\mu_{\tilde{2}}\;,
	& \psi_{2\tilde{2}} &=& s_{2}\mu_{\tilde{2}}\;.
\end{array}
\label{parafermions}
\end{equation}

\begin{figure}
\begin{center}
\includegraphics[angle=0,scale=0.5]{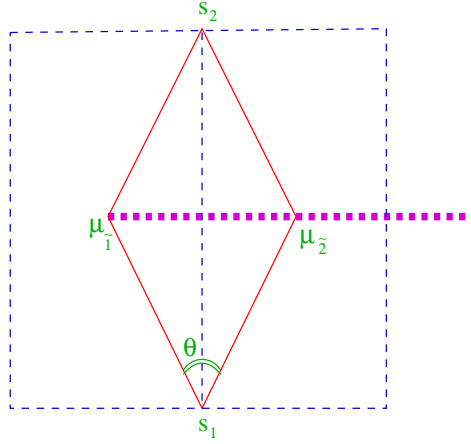}
\caption{The faces of the covering lattice form rombi with angle $\theta$; the thick dashed line is the frustration line.}
\end{center}
\end{figure} 

To give a geometrical meaning to $\theta$ consider the quadrilateral $(s_{1},\mu_{\tilde{1}},s_{2},\mu_{\tilde{2}})$
to be a rhombus with angle $\theta$ on the spin-variables vertices. Then one can write Eq.~(\ref{equation_of_motion_1}) as follows:
\begin{equation}\label{equation_of_motion_2}
\sum_{e}\psi_{e}\delta z_{e}=\Big(e^{i\theta}+\frac{w_{y}-i}{w_{y}+i}e^{i\frac{\theta}{2}}\Big)(\psi_{1\tilde{2}}-\psi_{2\tilde{1}})\;,
\hspace{0.5cm}\mbox{with}\hspace{0.5cm} \psi_{r\tilde{r}}=e^{-\frac{i\theta_{r\tilde{r}}}{2}}s_{r}\mu_{\tilde{r}}\;,
\end{equation}
where $\theta_{r\tilde{r}}$ is the angle that the directed segment $r\tilde{r}$ makes with the axis parallel to $s_{1}$-$\mu_{\tilde{1}}$.
The left-hand side of the above equation is akin to a discrete contour integral around elementary rhombi of the covering lattice. By replacing $i\rightarrow -i$
one can find a similar equation for the discrete antiholomorphic spinor-like variables, with the following property:
\begin{equation}\label{antiholomorphic_summation}
\bar{\psi}_{1\tilde{1}}+\bar{\psi}_{1\tilde{2}}+\bar{\psi}_{2\tilde{1}}+\bar{\psi}_{2\tilde{2}}=
\Big(e^{i\theta}+e^{i\frac{\theta}{2}}\frac{1-w^{2}_{y}}{1+w^{2}_{y}}+ie^{i\frac{\theta}{2}}\frac{2w_{y}}{1+w^{2}_{y}}\Big)
(\psi_{2\tilde{1}}-\psi_{1\tilde{2}})\;.
\end{equation}
Using the Eqs.~(\ref{equation_of_motion_2}) and (\ref{antiholomorphic_summation}) one obtains the following equation:
\begin{equation}\label{dirac_equation}
\sum_{e}\psi_{e}\delta z_{e}=
i\Big(2\sin\Big(\frac{\theta}{2}\Big)\frac{1-w^{2}_{y}}{1+w^{2}_{y}}-2\cos\Big(\frac{\theta}{2}\Big)\frac{2w_{y}}{1+w^{2}_{y}}\Big)\sum_{e}\bar{\psi}_{e}\;.
\end{equation}
The above relation is clearly similar to the equations of motion for a field theory of a free Majorana fermion, i.~e.~:
\begin{equation}\label{dirac_equation_continuum}
\partial_{\bar{z}}\chi=i\frac{m}{2}\bar{\chi}\;,
\end{equation}
where $\chi$ and $\bar{\chi}$ are the chiral components of the free Majorana fermion. The comparison then yields:
\begin{eqnarray}\label{discrete_continuum}
\chi &=& a\psi\;,\hspace{1cm}\bar{\chi}=b\bar{\psi}\;,\hspace{1cm}a=ib\;,\\
m &=& 8\sin\Big(\frac{\theta}{2}\Big)\frac{1-w^{2}_{y}}{1+w^{2}_{y}}-8\cos\Big(\frac{\theta}{2}\Big)\frac{2w_{y}}{1+w^{2}_{y}}\;.
\end{eqnarray}
It is easy to see that $m=0$ for $w_{y}=\tan(\frac{\theta}{4})$, and that $\psi$ is the discrete holomorphic operator. This means that this particular embedding of the covering lattice leads to a critical system. This gives a complete geometrical meaning to Eq.~(\ref{dirac_equation_continuum}).
The discrete Dirac equation for the Ising model was derived long ago by many authors, see for instance \cite{Dot}.
\begin{figure}
\begin{center}
\includegraphics[angle=0,scale=0.4]{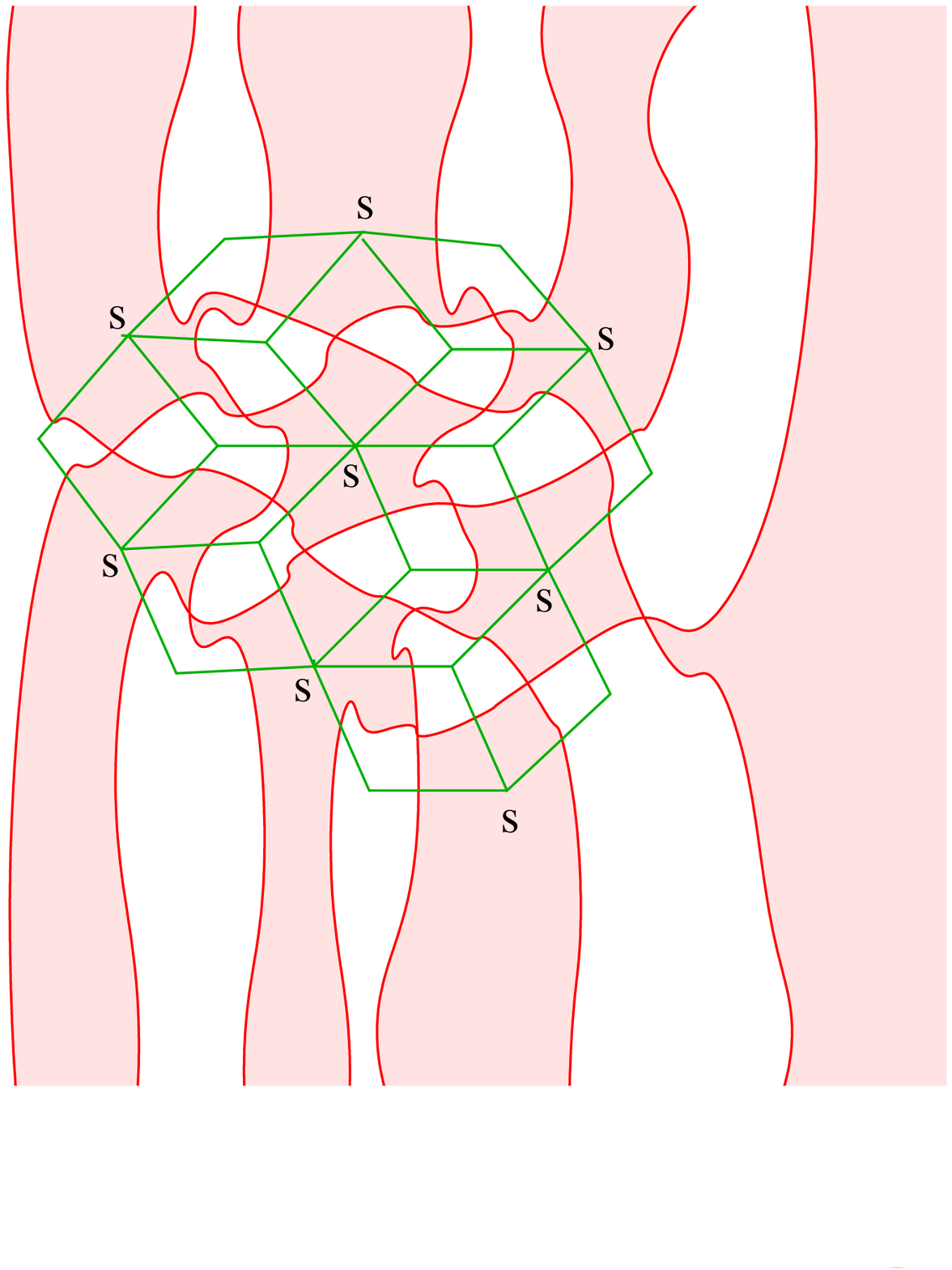}
\caption{2-colorable Baxter lattice and its Kenyon and Schlenker rombus embedding. We define the Ising variables in the shaded region 
and the disorder variables in the white region.}
\label{fig:circlepacking0}
\end{center}
\end{figure}

\begin{figure}
\begin{center}
\includegraphics[angle=0,scale=0.4]{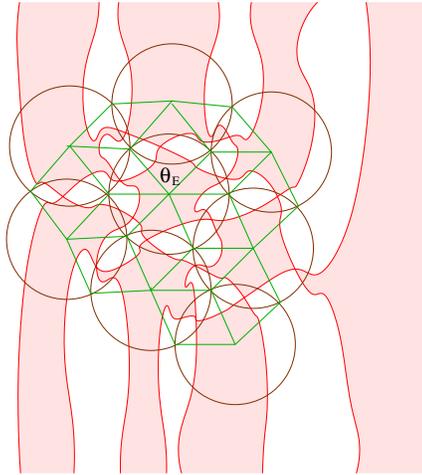}
\caption{The corresponding iso-radial packing of the  Kenyon and Schlenker rombus embedding. At the critical point $\theta _{E}$ is the only parameter 
needed to describe the critical temperature of the system.}
\label{fig:circlepacking}
\end{center}
\end{figure}

It is possible to extend the above results to the more general graphs called iso-radial graphs. To introduce iso-radial graphs we 
need first to define the $\mathbb{Z}$-invariant or Baxter lattice. This is a planar graph $\mathcal{L}$ which is a union of $M$ simple (non-self-intersecting) curves crossing the
complex plane from $x_{j}- i\infty$ to $x'_{j}+i\infty$, where the values ${x_{j}}$ and ${x'_{j}}$ ($1\leq j\leq M$) are distinct, and with the further
property that a given curve can intersect any of the others at most once (Fig.~\ref{fig:circlepacking0}). The faces of $\mathcal{L}$ are 2-colorable,
and in general we require that its vertices are all of degree four. Consider now the planar graph $\mathcal{G}$ whose vertices are associated 
to shaded faces of $\mathcal{L}$. We can define an Ising model on $\mathcal{G}$ with general weights $x^{E}_{k}$. The vertices of the dual lattice
$\mathcal{G}^{*}$ then correspond to the white faces of $\mathcal{L}$. The vertices of the covering graph $\mathcal{C}$ (the union of the vertices of $\mathcal{G}$ and
$\mathcal{G}^{*}$) correspond each to a face of $\mathcal{L}$, irrespective of color. There is a theorem, due to Kenyon and Schlenker \cite{KS}, stating that 
there is a rhombic embedding of $\mathcal{C}$ into the plane, that
is, one in which all the edges have equal length, see Fig.~\ref{fig:circlepacking0}. Each rhombus corresponds to an edge $E$ of $\mathcal{G}$, and defines an opening
angle $\theta_{E}$. This graph is called iso-radial graph because one can draw circles with equal radius centered on vertices of $\mathcal{G}$; the intersection between the circles will then define $\mathcal{G}^{*}$, see Fig.~\ref{fig:circlepacking}. These kinds of graphs are very important because they are the most general graphs that can offer 
a discrete complex analysis with a well-defined continuum limit. Most of the well-known planar graphs are just instances of iso-radial graphs.
For example, the triangular lattice is an iso-radial graph with $\theta=\frac{\pi}{3}$ and the honeycomb lattice has $\theta=\frac{2\pi}{3}$.
As before, we can define the Ising model and its dual on this graph and extract Eq.~(\ref{dirac_equation})
with the substitution $\theta\rightarrow \theta_{E}$. It is easy to see that one can also extract the critical point of 
the Ising model on iso-radial graphs by requiring:
\begin{equation}\label{critical_point_Ising}
w_{E}=\tan\Big(\frac{\theta_{E}}{4}\Big)\;.
\end{equation}
The above simple example shows that one can extract the critical line of some models
starting from the sole knowledge of the critical point for the
anisotropic square lattice. In the next Section we will use the same technique to extract the critical line of the Ashkin-Teller model 
on the general iso-radial graph. The above argument shows that the the critical and off-critical Ising model may have a well-defined continuum limit on iso-radial graphs. The continuum limit of the critical Ising model was already rigorously investigated in \cite{CS}, and the above argument indicates that possibly the same 
graphs give the right answer in the continuum limit for the off-critical case.

\section{\label{sec:Anisotropic_Ashkin-Teller}Ashkin-Teller model on the iso-radial graphs}
\setcounter{equation}{0}
In this Section we will first give the critical surface for the Ashkin-Teller model on the generic iso-radial graph and then 
we will study holomorphic variables of the model at some particular points.

\subsection{\label{subsec:critical_line_of_at}Critical line of the Ashkin-Teller model}
In this Section, using the same argument as in the previous one, we give the exact form of the (integrable part of the) critical surface of the Ashkin-Teller model on the iso-radial graphs.
The only formula we need for this purpose is the critical line of the anisotropic Ashkin-Teller model on the square lattice. It is given in \cite{SP}
and has the following expression:
\begin{eqnarray}\label{critical_line_anisotropic_AT}
\theta&=&\frac{\pi(\lambda-u)}{\lambda}\;,\\
\frac{\sin(u)}{\sin(\lambda)}&=&\tanh(2\beta_{x})\;,\\
\frac{\sin(\lambda-u)}{\sin(\lambda)}&=&\tanh(2\beta_{y})\;,\\
\frac{\Delta}{2}=\cos(\lambda)&=&\frac{\sinh(2\alpha_{x})}{\sinh(2\beta_{x})}=\frac{\sinh(2\alpha_{y})}{\sinh(2\beta_{y})}\;,
\end{eqnarray}
where $\theta$ is the anisotropy angle, $u$ is the spectral parameter and $\lambda$ is a parameter labelling the universality 
class\footnote{Specifically, we have $\lambda_{P}=0$, $\lambda_{FZ}=\frac{\pi}{4}$, $\lambda_{I}=\frac{\pi}{2}$, $\lambda_{T}=\frac{2}{3}\pi$ and $\lambda_{XY}=\frac{3}{4}\pi$.}. 
The above formulas give the two-dimensional exact solution manifold of the anisotropic Ashkin-Teller model among the four 
couplings. By ``exact solution'' we mean that they
satisfy the star-triangle relations on this manifold \cite{SP, BP}. This solution can not be the full critical surface of the Ashkin-Teller model, however, it
is the largest manifold satisfying the star-triangle relation.  
The parameter $\lambda$ has the following lattice-independent relation with the correlation exponent (as can be obtained from the square-lattice case):
\begin{equation}\label{correlation_exponent}
\nu(\lambda)=\frac{2\pi-2\lambda}{3\pi-4\lambda}\;.
\end{equation}

Now the only thing we need to do is considering $\theta$ as the angle of the rhombus in the iso-radial graph. Let's consider some examples:
for the square lattice we have $\theta=\frac{\pi}{2}$, which leads to the following critical line:
\begin{eqnarray}
\sinh (2\beta^S)e^{2\alpha^S}&=&1\;, \\
x_2^S &=& 1 - 2 x_1^S \;, \label{eq:critline_x1x2_square}
\end{eqnarray}
and, for the parameter $\lambda$:
\begin{equation}
\lambda_S = 2 \arccos \frac{1}{2 \tanh 2 \beta}\;.
\end{equation}

\begin{figure}
\begin{center}
\includegraphics[angle=0,scale=0.5]{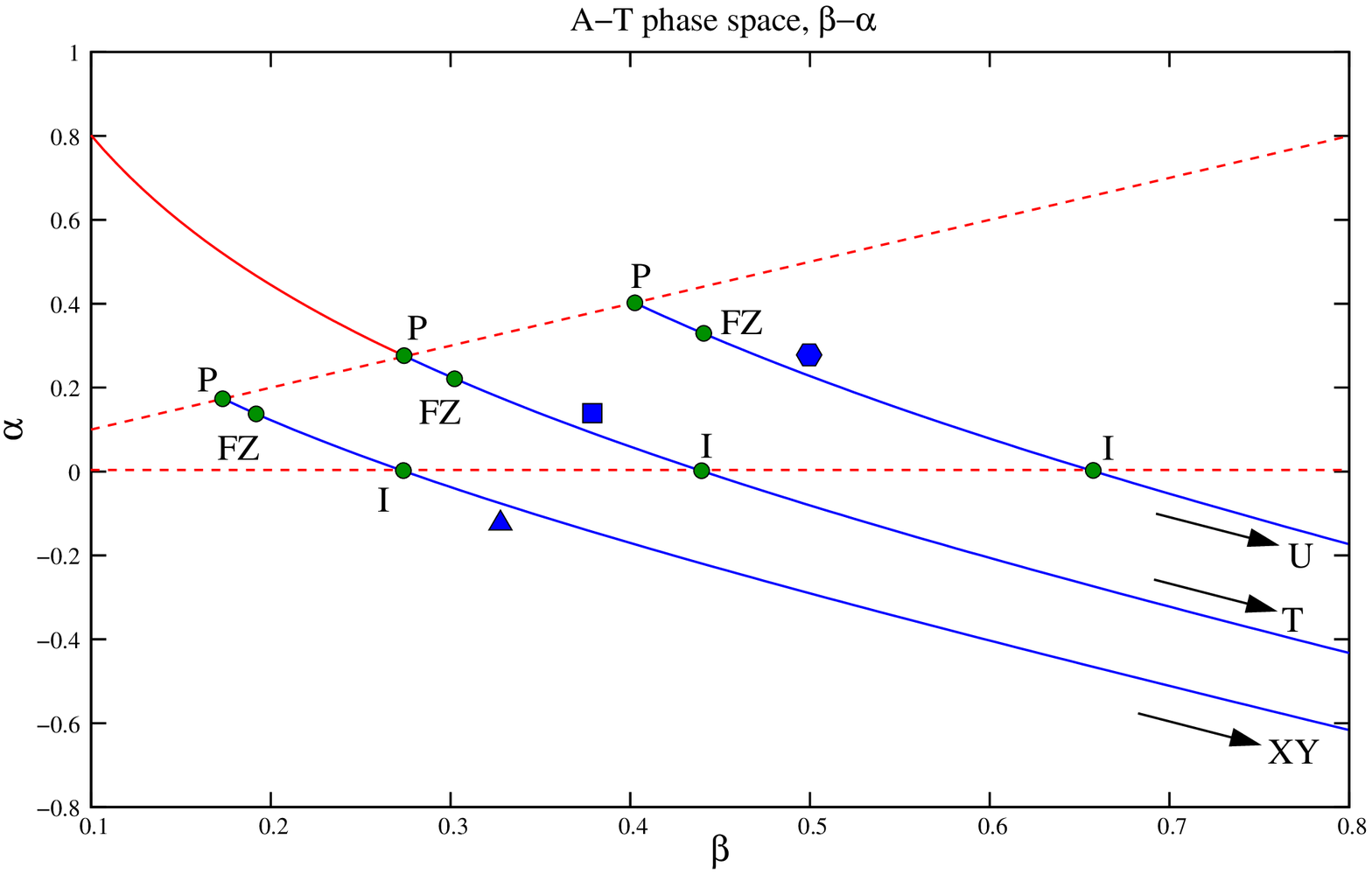} \\
\vspace{0.26cm}
\includegraphics[angle=0,scale=0.5]{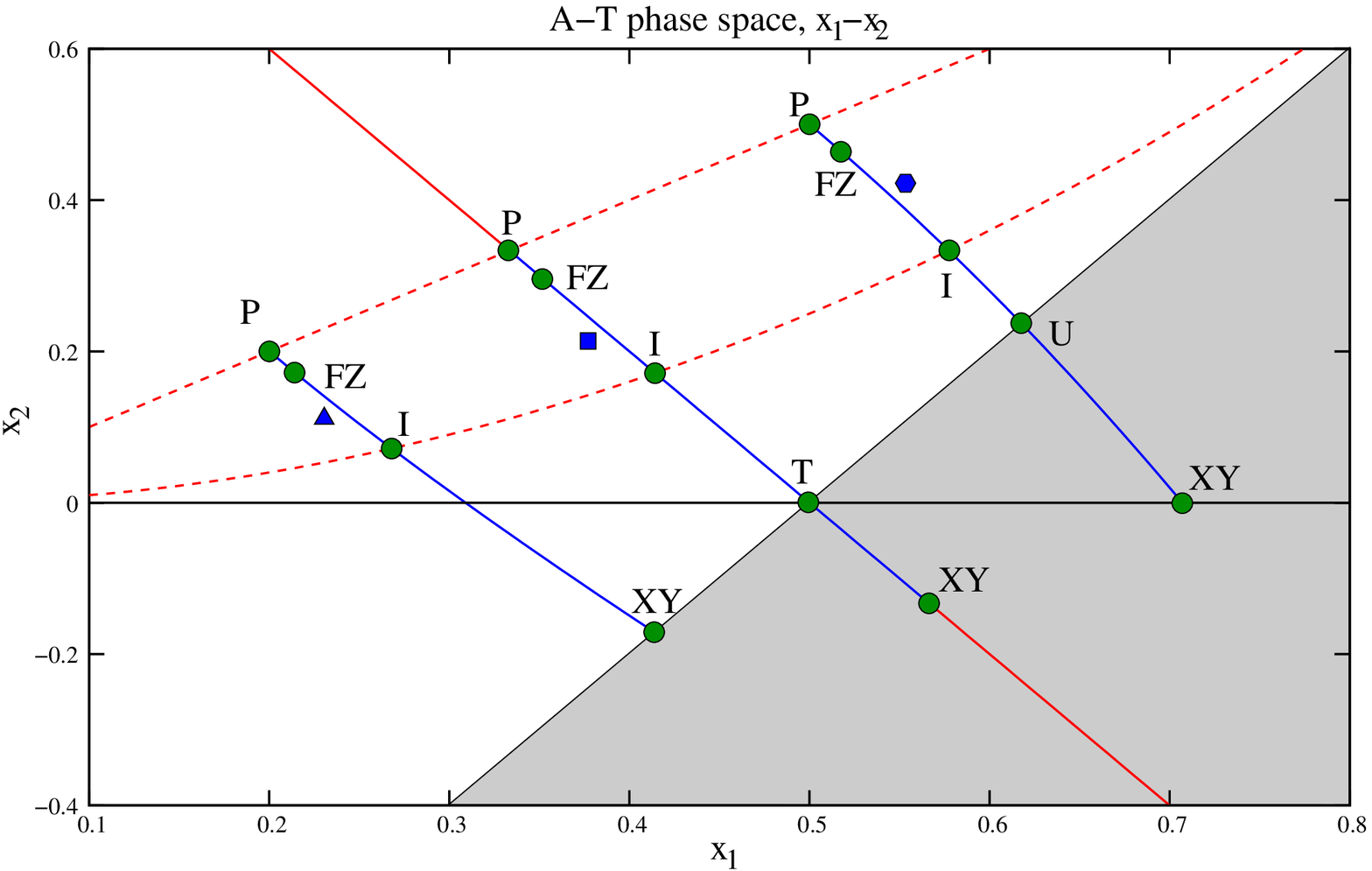}
\caption{
The critical lines for the Ashkin-Teller model on the triangular, square and honeycomb lattices (denoted by the symbols $\triangle$, $\square$ and $\hexagon$ respectively), depicted in the $\beta$-$\alpha$ (top) and the $x_1$-$x_2$ (bottom) planes. Dashed lines are the Potts and Ising lines; the critical lines span the interval between the Potts (P) and the $X$-$Y$ (XY) points; the self-dual line is also shown outside the (square-lattice) critical interval. In the $x_1$-$x_2$ plot, the shaded region is the non-physical one, i.~e.~the one having negative Boltzmann weights: as such, that region cannot be mapped to the $\beta$-$\alpha$ plane, and there the points on the threshold are mapped to infinity (they are the $X$-$Y$ point for triangular, the ``terminator'' $T$ point for square and a point labelled $U$ for the honeycomb lattice). For more information on the triangular- and honeycomb-lattice critical lines, see Appendix \ref{app:triang_honeycomb}.
}
\label{fig:at_critical_lines}
\end{center}
\end{figure}

To get the critical line 
of the triangular lattice we need to consider $\theta=\frac{\pi}{3}$; then we have the following lines in the $\beta$-$\alpha$ and $x_{1}$-$x_{2}$ planes:
\begin{eqnarray}
e^{4\alpha^{T}}(e^{4\beta^{T}}-1)&=&2\;,\\
x^{T}_{2}&=&1+x^{T}_{1}-\sqrt{5(x^{T}_{1})^{2}+4x^{T}_{1}}\;.\label{eq:critline_x1x2_triangular}
\end{eqnarray}
Here $\lambda$ has the following relation with the coupling constants:
\begin{equation}\label{triangular_critical_line_lambda_triangular}
\lambda_T=3 \arcsin \left(\sqrt{\frac{3\sinh 2\beta-\cosh 2\beta}{4\sinh 2\beta}}\right)\;.
\end{equation}
It is easy to see that in this case one can reach the $X$-$Y$ point $\nu\rightarrow \infty$ by staying in the physical region, i.~e.~with real $\beta$ and $\alpha$.
A similar calculation can be done for honeycomb lattice, with $\theta=\frac{2\pi}{3}$, finding:
\begin{eqnarray}\label{Honeycomb_critical_line}
e^{4\alpha^{H}}\sinh^{2}(2\beta^{H})-e^{2\alpha^{H}}\cosh(2\beta^{H})&=&1\;,\\
x^{H}_{2}&=&1-2(x^{H}_{1})^{2}\;.\label{eq:critline_x1x2_honeycomb}
\end{eqnarray}
In this case $\lambda$ can be written as:
\begin{equation}\label{triangular_critical_line_lambda_honeycomb}
\lambda_H=3 \arccos \left(\frac{\cosh(2\beta)+\sqrt{5\cosh^{2}2\beta-4}}{4\sinh 2\beta}\right)\;.
\end{equation}
On the honeycomb lattice, the universality classes falling in the physical region are a subset of those accessible in the triangular and square lattices (for a comparision of the three cases, see Fig.~\ref{fig:at_critical_lines}).
The same conclusions can be extracted by using the duality relations, Eqs.~(\ref{duality1}) and (\ref{duality2}). The above results are consistent with those in \cite{TA}.\footnote{Note that some parts of that paper, identifying the critical line of the Ashkin-Teller model with the universality class of the Potts model, are wrong.}

The same procedure is also effective for the anisotropic triangular lattice with different weights $\beta_{i}$ and $\alpha_{i}$ on the triangle sides $i=1,2,3$. Consider a (proper) embedding 
of the regular triangular lattice into the plane by some general linear transformation of the coordinates. Then locate the dual vertex at the circumcenter, 
that is, the point at which the three perpendicular bisectors of the edges meet, equidistant from the three vertices. This construction guarantees 
that adjacent pairs of vertices always form a rhombus. Each triangle is associated to three different rhombi of angles 
$\theta_{1}$, $\theta_{2}$ and $\theta_{3}=\pi-\theta_{1}-\theta_{2}$. Finally, one can write the critical line as follows:
\begin{eqnarray}
\theta_{i}&=&\frac{\pi(\lambda-u_{i})}{\lambda}\;,\label{critical_line_anisotropic_AT_on_the_anisotropic_triangular_1} \\
\frac{\sin(\lambda-u_{i})}{\sin(\lambda)}&=&\tanh(2\beta_{i})\;,\label{critical_line_anisotropic_AT_on_the_anisotropic_triangular_2} \\
\frac{\Delta}{2}=\cos(\lambda)&=&\frac{\sinh(2\alpha_{i})}{\sinh(2\beta_{i})}\;,\label{critical_line_anisotropic_AT_on_the_anisotropic_triangular_3}
\end{eqnarray}
with $i=1,2,3$. It is not difficult to see that the above formulas give a three-dimensional manifold among the six thermodynamic parameters.
This can not be the full critical manifold of the system. The point is, our method of extracting the critical manifold is in close relation with the 
integrability of the system. On this manifold the star-triangle relations among the couplings are satisfied, meaning that our method works only as long as the star-triangle relations are satisfied; the same observation was also made for the Fateev-Zamolodchikov point in the lattice $\mathbb{Z}_N$ models \cite{RC}. The close connection of our method to the star-triangle relations can be related to the integrability 
properties of the iso-radial graphs and can come from deep geometrical reasons, see \cite{BMS} and references therein. The above formulas are also true for the anisotropic honeycomb lattice. The only subtlety is, we need to consider three angles 
$\theta_{1}$, $\theta_{2}$ and $\theta_{3}$ constrained by $\theta_{1}+\theta_{2}+\theta_{3}=2\pi$.

\subsection{\label{sec:free_energy}Free energy}
In this Subsection we write the exact free energy of the Ashkin-Teller model on the critical surface of iso-radial graphs. To avoid any non-positive 
Boltzman weight we will concentrate on the region between the Ising and the 4-state Potts models.
To find the free energy on iso-radial graphs we first need to find the free energy on the anisotropic square lattice; in order to do so, we need the unitarity conditions \cite{Zamol}. We follow the notation of \cite{SP} to keep the equations simple. Consider the weights 
\begin{eqnarray}\label{AT_new_weights}
W_{x}(i,j) &=& \rho_{x} \exp[\beta_{x}(\sigma_{i}\sigma_{j}+\tau_{i}\tau_{j})+\alpha_{x} (\sigma_{i}\sigma_{j}\tau_{i}\tau_{j})]\;;\\
W_{y}(i,j) &=& \rho_{y} \exp[\beta_{y}(\sigma_{i}\sigma_{j}+\tau_{i}\tau_{j})+\alpha_{y} (\sigma_{i}\sigma_{j}\tau_{i}\tau_{j})]\;,
\end{eqnarray} 
where $\rho_{x}$ and $\rho_{y}$ are some normalisation constants, that can take any finite value and only shift the free energy by a constant.
It is convenient to use the following parametrisation:
\begin{equation}\label{parametrization}
s=\frac{\sin(u)}{\sin(\lambda)}\;, \hspace{0.8cm} s_{-}=\frac{\sin(\lambda-u)}{\sin(\lambda)}\;, \hspace{0.8cm}s_{+}=\frac{\sin(\lambda+u)}{\sin(\lambda)}\;.
\end{equation}
If we choose $\rho_{x,y}=\frac{1}{\sqrt{2}}e^{\alpha_{x,y}}\tanh 2\beta_{x,y}$, the unitarity conditions, or inversion identities, will be:
\begin{eqnarray}\label{unitarity}
\sum_{k}W_{x}(i,k|u)W_{x}(k,j|-u) &=& 2s_{+}s_{-}\delta(i,j)\;,\\
W_{y}(i,j|u)W_{y}(i,j|-u) &=& \frac{1}{2}s_{+}s_{-}\;.
\end{eqnarray}
Following \cite{baxter2,Zamol}, one can argue that the free energy should satisfy:
\begin{equation}\label{free_energy_unitarity}
f(\theta)+f(\pi+\theta)=\log(2s_{+}s_{-})\;, \hspace{0.8cm} f(\theta)=f(\pi-\theta)\;.
\end{equation}
The above equations were solved in \cite{BP}; one can write the free energy of the Ashkin-Teller model on the anisotropic square lattice as follows:
\begin{equation}\label{free_energy_square_lattice}
f(\theta,\lambda)=\log(ss_{-})+2\int_{0}^{\infty}\frac{\sinh[(\pi-\lambda)x]}{\sinh(\pi x)}\tanh(\lambda x)\cosh\Big[\Big(1-2\frac{\theta}{\pi}\Big)\lambda x\Big]\frac{dx}{x}\;.
\end{equation}
For the free energy of the system on the iso-radial graphs,\footnote{The free energy of iso-radial graphs is the same as in Baxter lattices.} consider a graph with a very large number of bulk edges $N_{E}$, so that the number of exterior sites scales as $\sqrt{N_{E}}$: then, by following Section 5 of \cite{baxter2}, one can show that the free energy for the iso-radial graphs is simply the sum of the per-site free energies: in other words,
\begin{equation}\label{free_energy}
f=\sum_{(ij\in E(\mathcal{G}))}f(\theta_{ij})+O(\sqrt{N_{E}})\;,
\end{equation}
where the function $f(\theta_{ij})$ is the same as in Eq.~(\ref{free_energy_square_lattice}), and the summation runs over all edges of the graph $\mathcal{G}$. It is worth mentioning here that to get the free energy of the system on iso-radial graphs it is sufficient to know the free energy of the system on a particular lattice, not necessarily square; in other words, $f(\theta_{ij})$ does not depend on the specific details of the lattice.

\subsection{\label{sec:holomorphic_variables}Holomorphic variables at some particular points}
In this Section we want to show that it is possible to extract some holomorphic variables, written with respect to
spin and disorder variables, for the Ashkin-Teller model at some particular points on the critical line. The holomorphic variables can be written either in terms of $\sigma$ and $\tau$ or $s$ variables. The definitions of disorder variables with respect to the $\sigma$ and $\tau$ spin variables are:
\begin{eqnarray}\label{disorderAT1}
\mu_{\sigma\tilde{1}}&=&\frac{1+x_{1}(-\sigma_{1}\sigma_{2}+\tau_{1}\tau_{2})}{1+x_{1}(\sigma_{1}\sigma_{2}+\tau_{1}\tau_{2})}\mu_{\sigma\tilde{2}}\;, \\
\mu_{\tau\tilde{1}}&=&\frac{1+x_{1}(\sigma_{1}\sigma_{2}-\tau_{1}\tau_{2})}{1+x_{1}(\sigma_{1}\sigma_{2}+\tau_{1}\tau_{2})}\mu_{\tau\tilde{2}}\;.
\end{eqnarray} 
The disorder variable with respect to the complex spin variable $s$ is given by:
\begin{equation}\label{disorderAT2}
\mu_{s1}=\frac{1+x_{1}(\omega^{*} s_{1}s^{*}_{2}+\omega s^{*}_{1}s_{2})-\omega^{2}x_{2}(s^{*2}_{1}s^{2}_{2})}{1+x_{1}(s_{1}s^{*}_{2}+
s^{*}_{1}s_{2})+x_{2}(s^{*2}_{1}s^{2}_{2})}\mu_{s2}\;,
\end{equation} 
where $\omega^{4}=1$. There is no simple formula directly relating $\mu_{s}$ and $(\mu_{\sigma},\mu_{\tau})$.
From the first two equations one can get, for the Ising point, the following two parafermionic 
operators on the generic iso-radial graphs:
\begin{equation}\label{holomorphic1}
\psi_{\sigma r\tilde{r}}=e^{-\frac{i\theta_{r\tilde{r}}}{2}}\sigma_{r}\mu_{\sigma\tilde{r}}\;,\hspace{1cm}\psi_{\tau r\tilde{r}}=e^{-\frac{i\theta_{r\tilde{r}}}{2}}\tau_{r}\mu_{\tau\tilde{r}}\;,
\end{equation}
where, again, $\theta_{r\tilde{r}}$ is the angle the directed segment $r\tilde{r}$ 
makes with the axes parallel to $\sigma_{1}$-$\mu_{\sigma \tilde{1}}$ or $\tau_{1}$-$\mu_{\tilde{1}}$. It is easy also to show that at 
the 4-state Potts point the following variables satisfy the discrete Cauchy-Riemann equations on the iso-radial graphs:
\begin{equation}\label{holomorphic2}
\psi_{\sigma r\tilde{r}}=e^{-i\theta_{r\tilde{r}}}\sigma_{r}\mu_{\tau\tilde{r}}\;,\hspace{1cm}\psi_{\tau r\tilde{r}}=e^{-i\theta_{r\tilde{r}}}\tau_{r}\mu_{\sigma\tilde{r}}\;.
\end{equation}
The above discrete variables could be  candidates for the continuum counterparts with spin $p=1$. Using the variables $s$ and $\mu_{s2}$ one can show that the following quantity satisfies, at the F-Z point, the discrete Cauchy-Riemann equations on the iso-radial graph: \cite{RC}
\begin{equation}\label{holomorphic3}
\psi_{s r\tilde{r}}=e^{-i\frac{3\theta_{r\tilde{r}}}{4}}s^{3}_{r}\mu_{\tilde{r}}\;.
\end{equation}
The above holomorphic variable is the discrete version of the parafermionic operator of Ref.~\cite{ZF}.
The next interesting holomorphic variable is related to the $X$-$Y$ point and has the following form:
\begin{equation}\label{holomorphic4}
\psi_{s r\tilde{r}}=e^{-i\frac{\theta_{r\tilde{r}}}{4}}s^{3}_{r}\mu_{\tilde{r}}\;.
\end{equation}
The above variable can be the discrete version of the parafermionic operator of spin $p=\frac{1}{4}$ in the continuum limit. 
From the continuum limit of the critical line of the Ashkin-Teller model, we know there should be some holomorphic operators all the way along the critical line \cite{Kadanoff_and_Brown}.
The spin of these holomorphic operators is given by $p=\frac{r}{4}$, with $r$ given by Eq.~(\ref{AT_R}). The above four operators
are consistent with this continuum prediction; however, it seems that at points other than those four it is not easy to write the 
holomorphic variables as a function of the discrete spin variables.

\section{\label{sec:montecarlo_triangular} Numerical results for the anisotropic triangular lattice}
This Section presents Monte Carlo results in support of the formula for the critical manifold, Eqs.~\ref{critical_line_anisotropic_AT_on_the_anisotropic_triangular_1}-\ref{critical_line_anisotropic_AT_on_the_anisotropic_triangular_3}, in the particular case of a triangular lattice with couplings $\beta_1,\alpha_1$ on two sides of the triangle and $\beta_2,\alpha_2$ on the other. Note that, in the final formulation of the model (as well as in the simulation algorithm used), only the values of the couplings on the different links encode the geometrical anisotropy and carry the information about the opening angle of the rhombi.

On such a system the angles are constrained by $2\theta_1+\theta_2 = \pi$; in the four-dimensional coupling space, the full critical line is expected to be three-dimensional, while its integrable part (in the sense specified earlier) is a two-dimensional surface.
After some manipulation of the seven starting relations (the constraint on the angles, plus Eqs.~\ref{critical_line_anisotropic_AT_on_the_anisotropic_triangular_1}-\ref{critical_line_anisotropic_AT_on_the_anisotropic_triangular_3} taken for $i=1$ and $i=2$), one can express the critical surface $\Sigma$ as explicitly parametrised by the two couplings $\beta_1,\alpha_1$:
\begin{eqnarray}
	\beta_2 &=& \frac{1}{2}\arctanh\Big( \frac{1-\sinh^22\beta_1+2\sinh^22\alpha_1-2\sinh2\alpha_1\cosh2\alpha_1}{\cosh^22\beta_1} \Big) \equiv \nonumber \\
		&&\equiv G(\beta_1,\alpha_1)\;; \label{eq:happytree_expl_1} \\
	\alpha_2 &=& \frac{1}{2} \arcsinh \Big( \frac{\sinh 2G(\beta_1,\alpha_1) \sinh 2\alpha_1}{\sinh2\beta_1} \Big)\equiv\nonumber\\
	&&\equiv H(\beta_1,\alpha_1) \;. \label{eq:happytree_expl_2}
\end{eqnarray}
This surface is the union of one-dimensional ``iso-$\lambda$'' lines, where the universality class stays constant; the universality class label is still given by Eq.~\ref{critical_line_anisotropic_AT_on_the_anisotropic_triangular_3}.

In order to confirm the validity of the above formulas, we chose four highly anisotropic points in the critical region between the four-state Potts and the Ising ``iso-$\lambda$'' lines and sampled the (density of) energy susceptibility along segments crossing the critical manifold. Then, the scaling behaviour of this quantity can be checked against the $\nu(\lambda)$ predicted from Eqs.~\ref{critical_line_anisotropic_AT_on_the_anisotropic_triangular_3} and \ref{correlation_exponent}. Such a signal is maximised if the segments are chosen to lie orthogonal to the critical surface, whose orientation can be found analytically from the parametrisation above (see Appendix \ref{app:anisotriang_gradients}). In four dimensions, the space orthogonal to a 2-D manifold is two-dimensional: thus we sampled the susceptibility along two directions $w_1$ and $w_2$ for each point, such that $w_1 \perp w_2$ and $w_i \perp \Sigma$.

There is still an important caveat to keep in mind: while the above formulas provide the exact critical manifold, i.~e.~critical \textit{and} integrable (satisfying the star-triangle relations), we know the complete critical manifold is here $3$-dimensional. It is then to be expected that a particular direction in the plane spanned by $(w_1,w_2)$ stays on criticality, but the chances of it being exactly one or the other of the two basis vector employed is vanishing; indeed, it turned out not to be the case for our numerical investigation. On the other hand, a direct trial-and-error numerical search for the orientation of the full critical manifold appears hardly feasible, time-consuming and of scarce conceptual relevance, so it has not been pursued further.

The four points chosen for the numerical investigation, along with the corresponding prediction for correlation index, are listed in Table \ref{tab:anisotriang_points}. 

\begin{table}
\begin{center}
\begin{tabular}{|c|l|l|l|l|c|}
\hline
Point name & $\beta_1$ & $\alpha_1$ & $\beta_2$ & $\alpha_2$ & expected $\nu$ \\
\hline
\hline
A1 & $0.205000$ & $0.202493$ & $0.115419$ & $0.113956$ & $0.678871$ \\
\hline
AFZ & $0.250000$ & $0.180302$ & $0.085211$ & $0.060398$ & $3/4$ \\
\hline
A2 & $0.277140$ & $0.120000$ & $0.105562$ & $0.044136$ & $0.823488$ \\
\hline
A3 & $0.170752$ & $0.035000$ & $0.402875$ & $0.089642$ & $0.897473$ \\
\hline
\end{tabular}
\end{center}
\caption{The four points on which the numerical results were collected for the anisotropic triangular lattice. The expected correlation exponent is also reported.}
\label{tab:anisotriang_points}
\end{table}

The (density of) energy susceptibility is defined as:
\begin{equation}
	U_L(\beta_1) = \Big\langle \Big(\Box-\langle\Box\rangle_{\beta_1;L}\Big)^2 \Big\rangle_{\beta_1;L}\;,
\end{equation}
where
\begin{equation}
	\Box = -\frac{1}{3L^2}\sum_{i=1}^2\sum_{\langle x,y \rangle_{i}}\Big( \beta_i(\sigma_x\sigma_y+\tau_x\tau_y) + \alpha_i(\sigma_x\sigma_y\tau_x\tau_y) \Big)\;,
\end{equation}
that is, the average energy per link in the lattice according to the action for the system (the neighbouring pairs of sites $\langle x,y\rangle_i$ are taken according to the corresponding anisotropic coupling). $L$ is the side, in lattice sites, of the system, which has periodic boundary conditions and a square shape. Measurements are collected for different values of $L$ and different points on the segments, and according to the predictions for a second-order transition the curves for various system sizes should collapse one onto another if one plots
\begin{equation}
	F = L^{4-\frac{2}{\nu}}U_L(\beta) \;\;\mbox{as a function of}\;\; x = (\beta_1-\beta_1^c)L^{\frac{1}{\nu}}\;,
	\label{eq:anisotriang_susc_rescale}
\end{equation}
with $\beta_1^c$ denoting criticality (obviously, any other coupling would work).

In this numerical investigation, 36 points were sampled along each of the two segments for the four points $A1$, $AFZ$, $A2$ and $A3$; the collected statistics was of 36000 configurations per each data point, with six system sides ranging from $L=150$ to $L=500$. The algorithm used was an adaptation to the anisotropic Ashkin-Teller model of the standard cluster-based Swendsen-Wang prescription, based on identifying and flipping the Fortuin-Kasteleyn clusters alternatively in the $\sigma$ and $\tau$ spin variables (with the other Ising spin playing the r\^ole of a local shift of the link coupling $\beta_i$ to $\beta_i\pm\alpha_i$); for more details on the algorithm, we refer to \cite{z4_gauge_dualAT}.

\begin{figure}
\begin{center}
	\includegraphics[height=6.5cm,angle=-90]{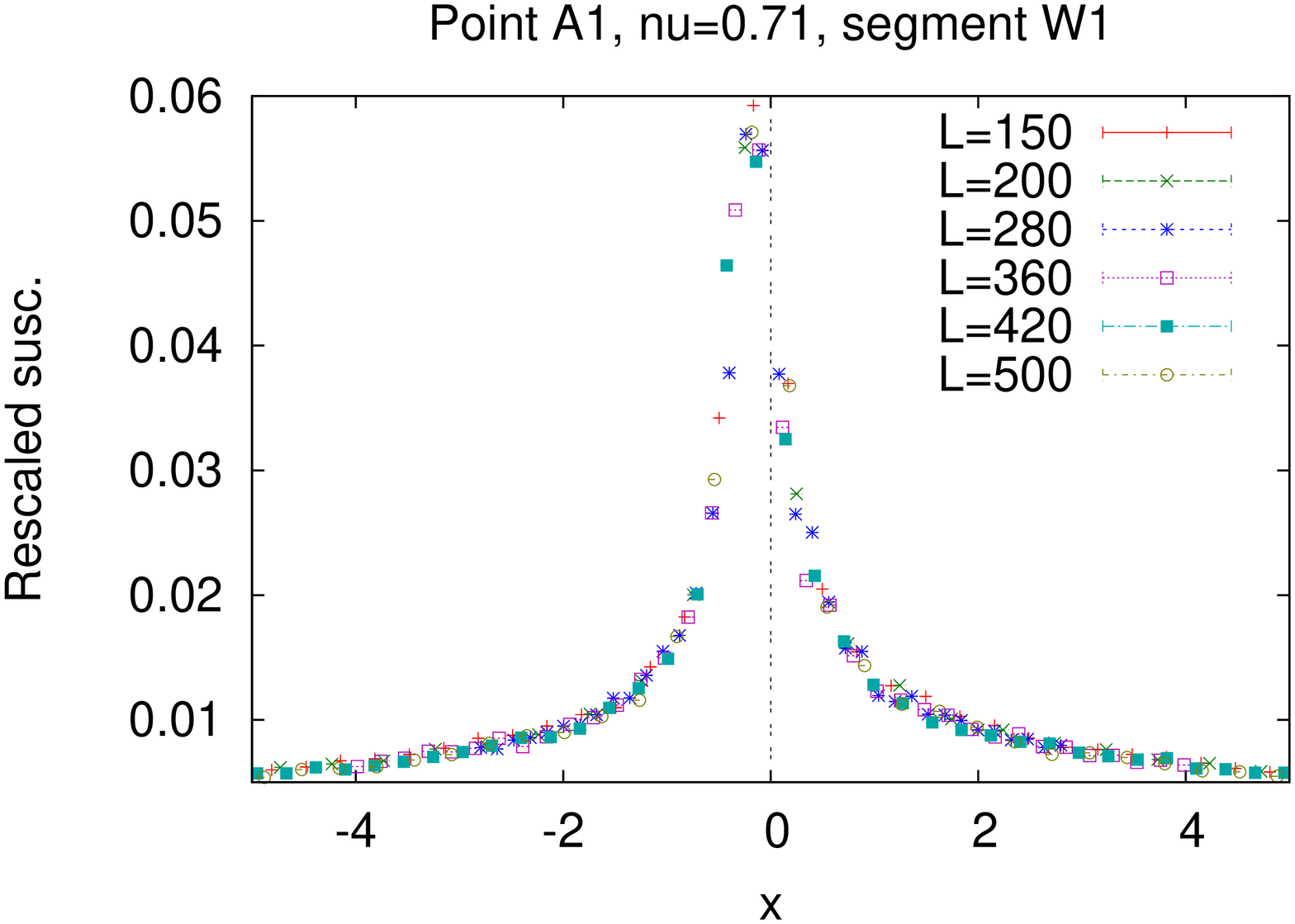}
		\hspace{1em}
	\includegraphics[height=6.5cm,angle=-90]{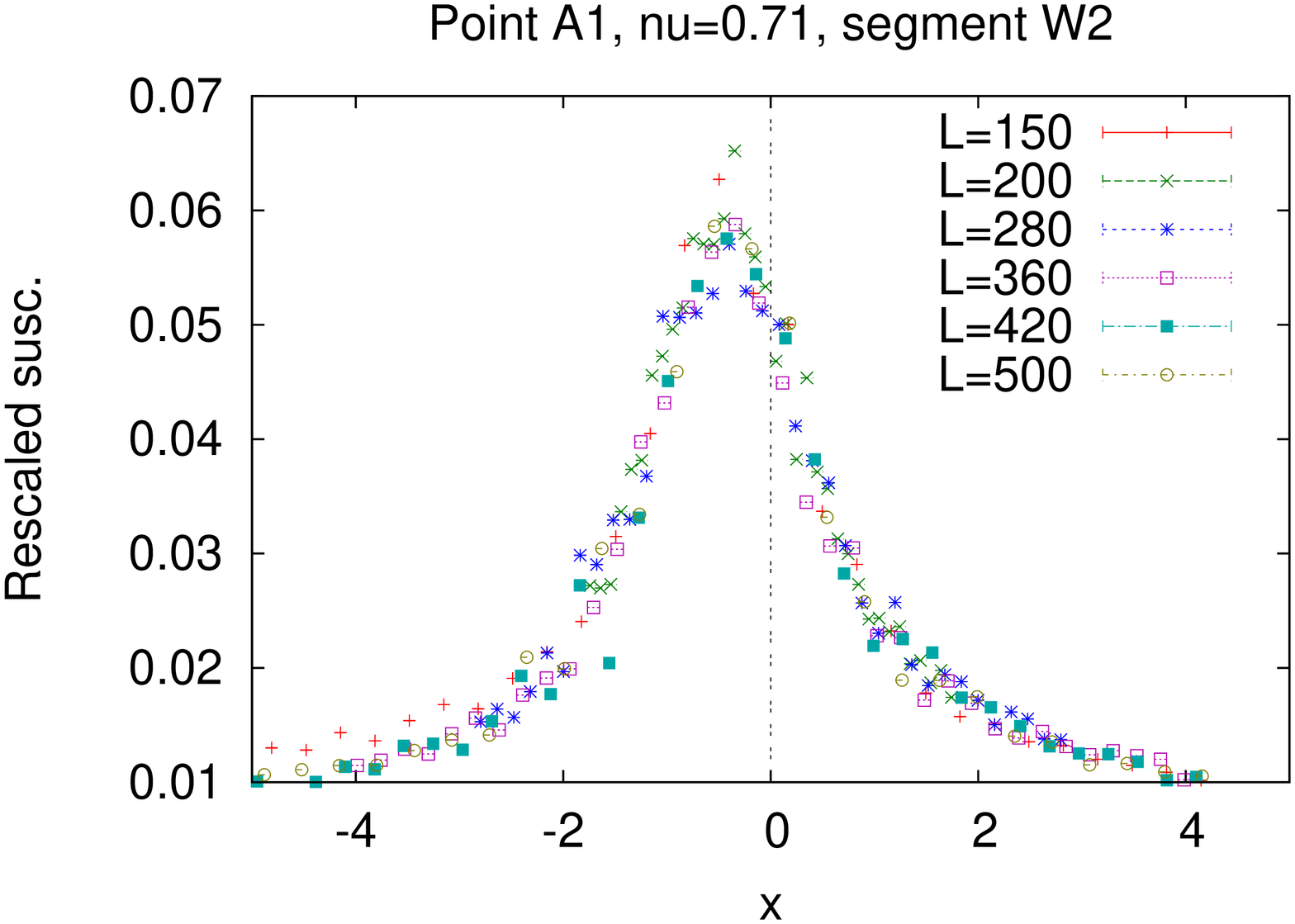} \\
	\includegraphics[height=6.5cm,angle=-90]{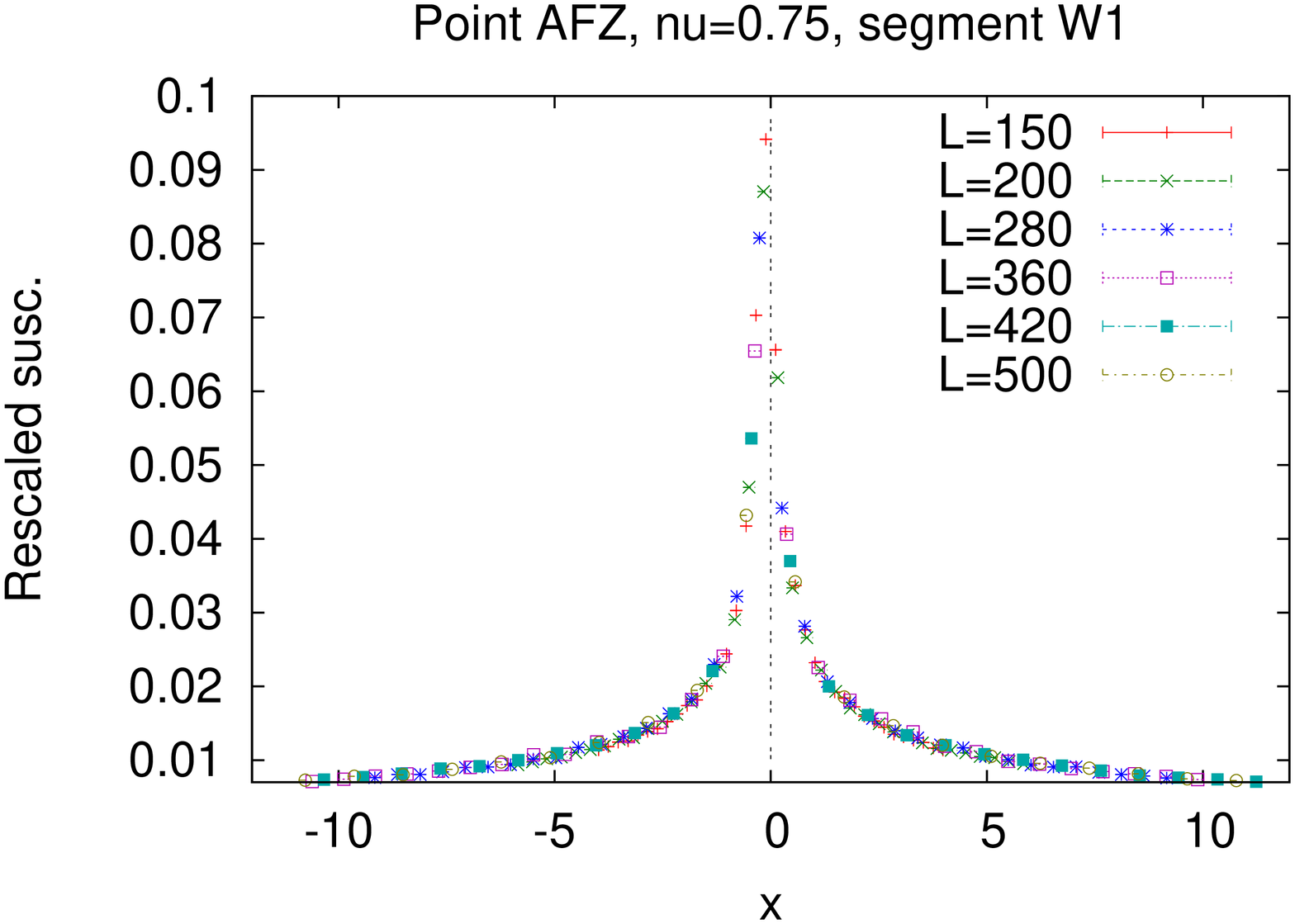}
		\hspace{1em}
	\includegraphics[height=6.5cm,angle=-90]{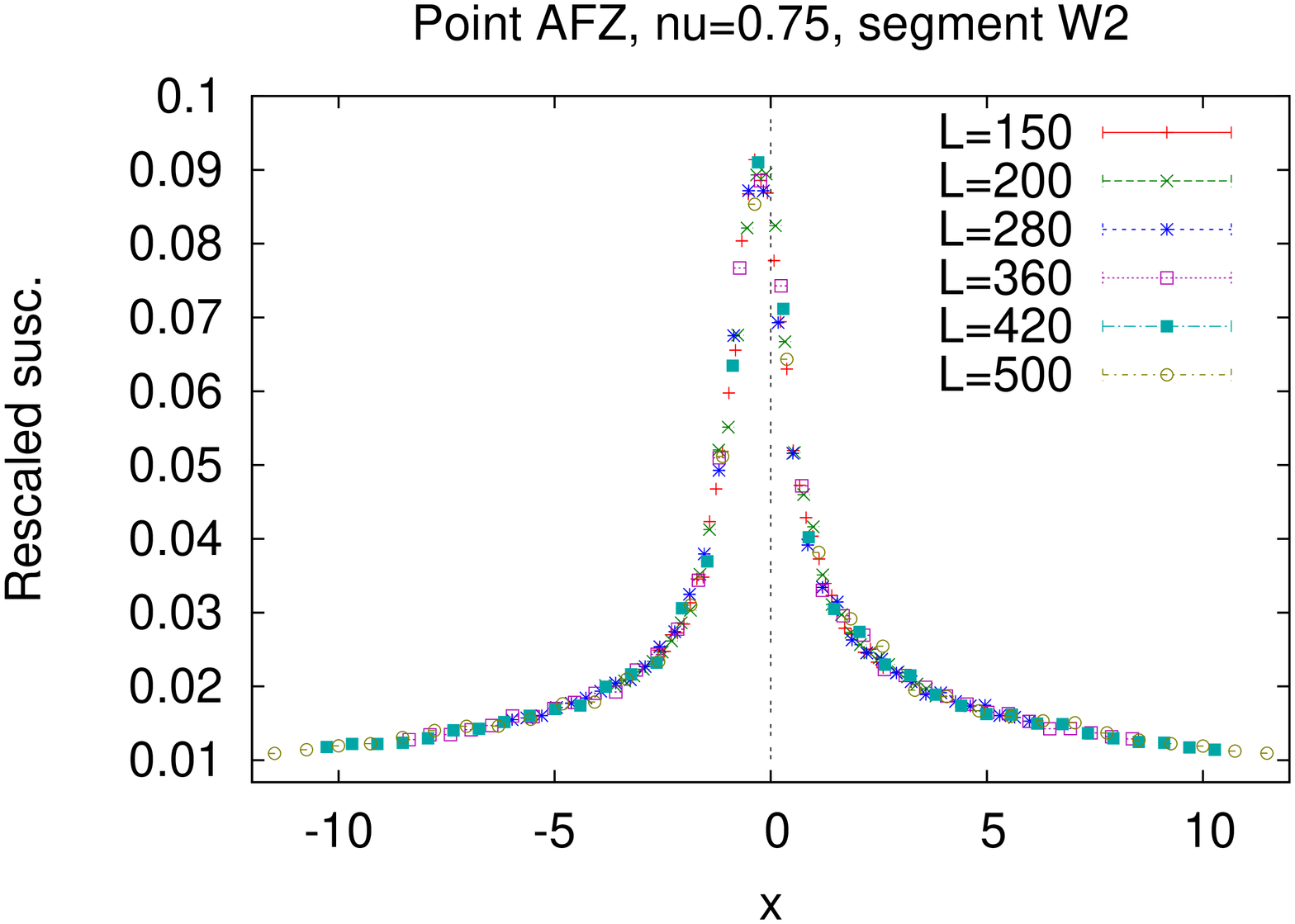} \\
	\includegraphics[height=6.5cm,angle=-90]{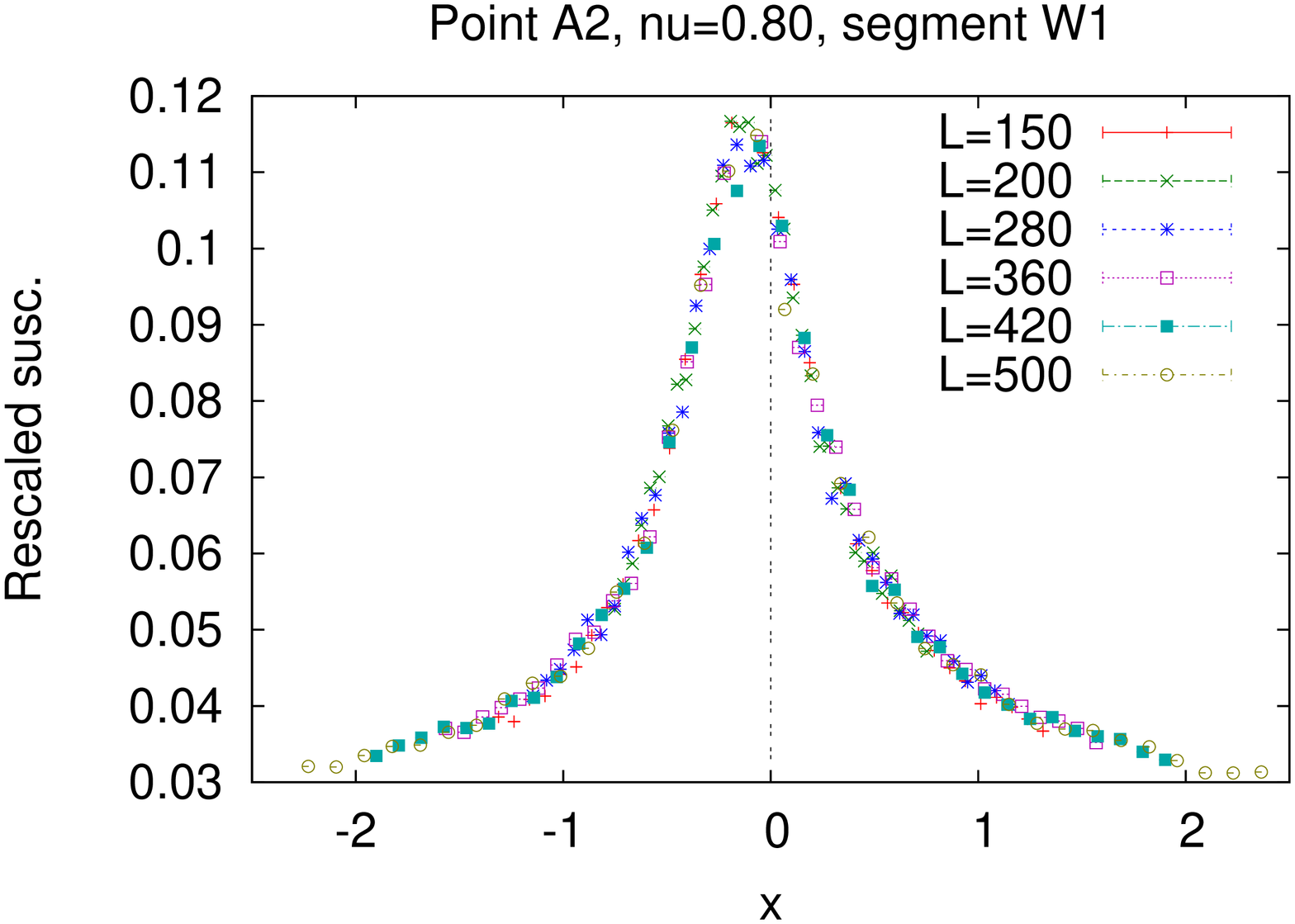}
		\hspace{1em}
	\includegraphics[height=6.5cm,angle=-90]{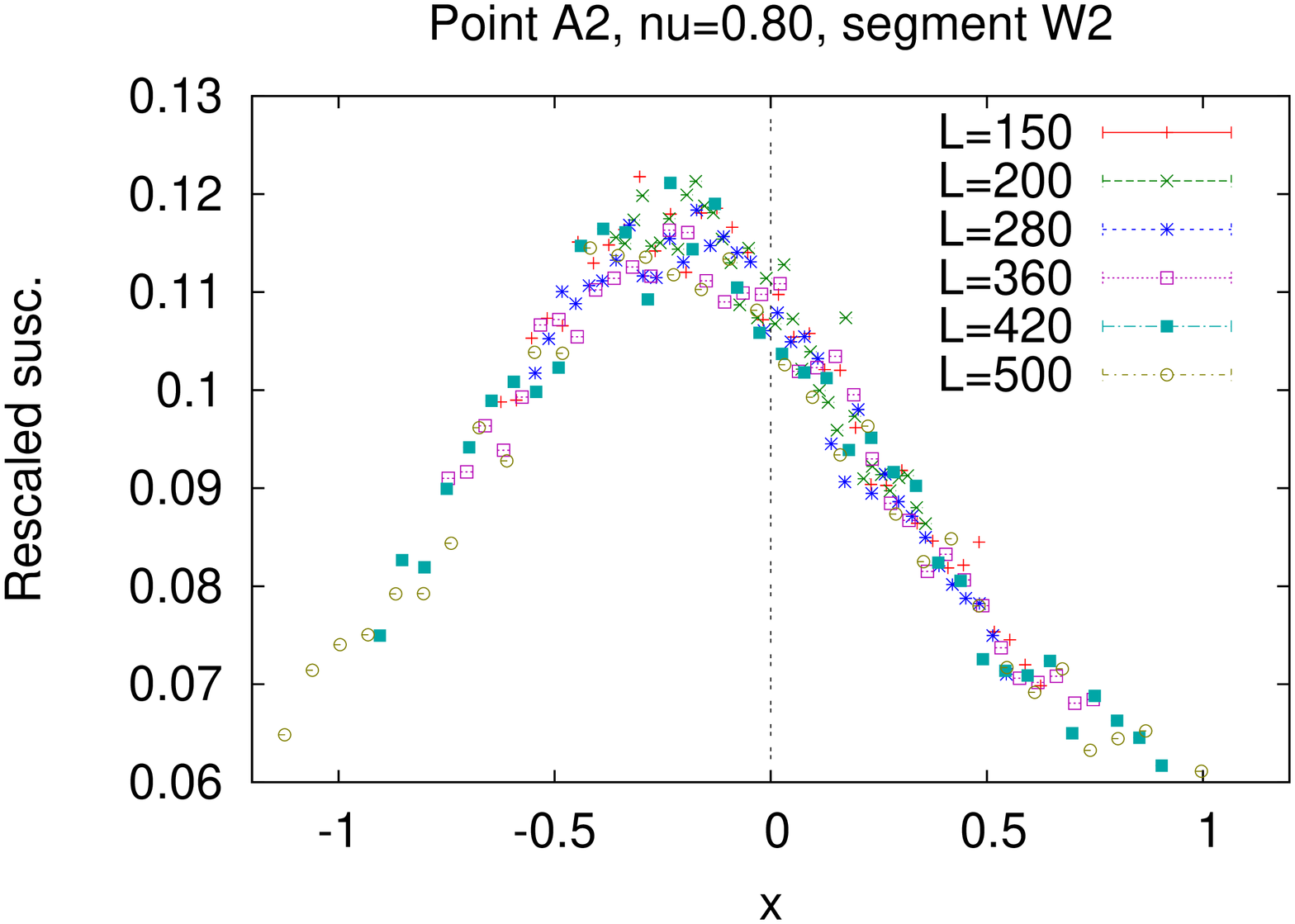} \\
	\includegraphics[height=6.5cm,angle=-90]{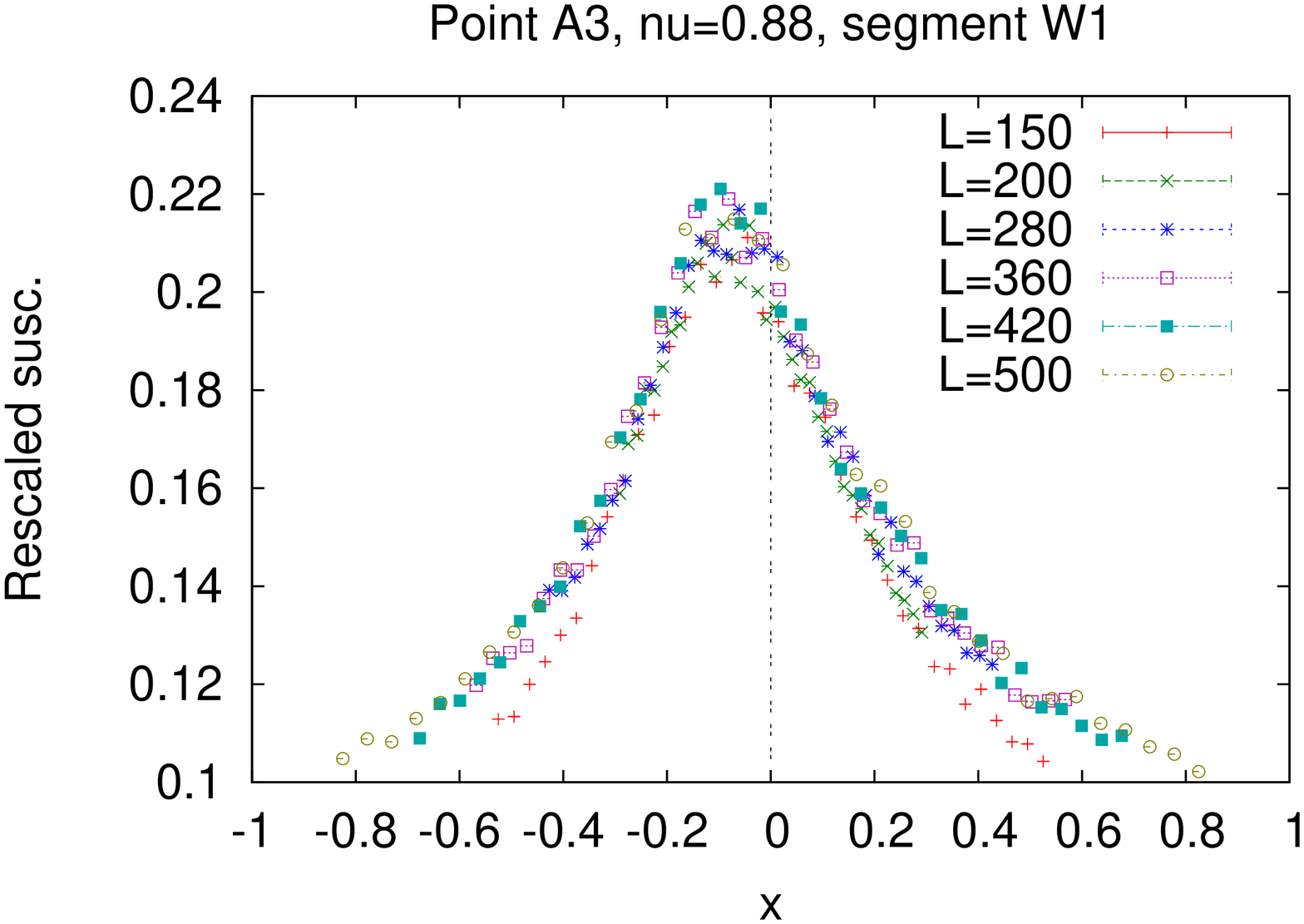}
		\hspace{1em}
	\includegraphics[height=6.5cm,angle=-90]{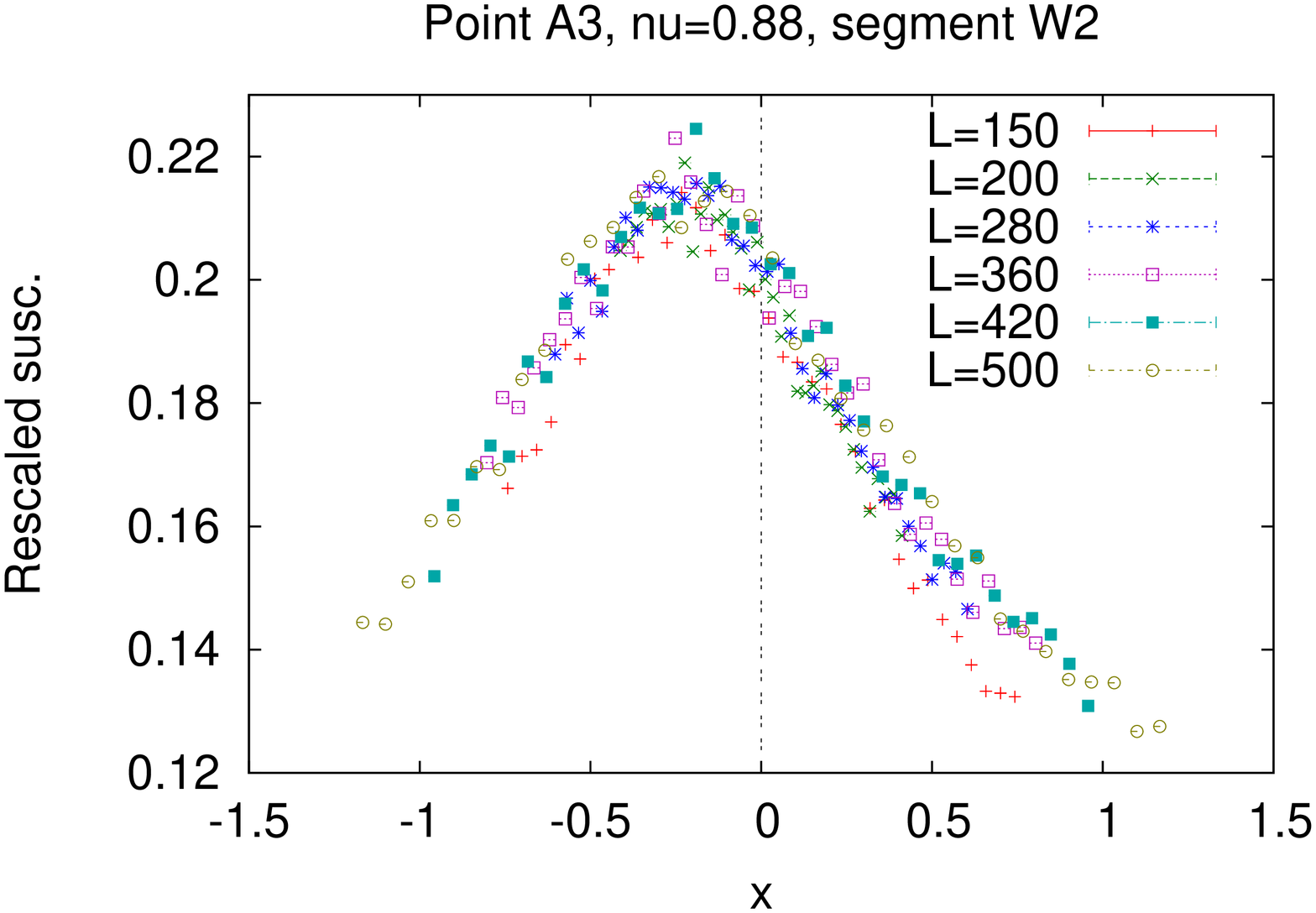}
	\caption{The susceptibilities for points $A1$, $AFZ$, $A2$ and $A3$ (top to bottom) along the two segments (left: $w_1$; right: $w_2$) as described in the text, rescaled according to Eq.~\ref{eq:anisotriang_susc_rescale}. The plots employ the ``experimental'' correlation index $\nu^*$ as indicated in the figures' caption.}
	\label{fig:anisotriang_susc_rescale}
\end{center}
\end{figure}

The rescaled susceptibilities fall wery well to a universal curve $F(x)$ in all four cases, thus confirming the location of the critical point along the segment and the accuracy of the prediction for the correlation index; to estimate the effect of systematic deviations on the latter (coming for instance from the finiteness of the system under study, or from the interplay between the anisotropy and the always square shape of the system), we compared the theoretical $\nu$ with a value $\nu^*$ tuned as to yield the best collapse of the data, finding a discrepancy never exceeding about 5\% (Fig.~\ref{fig:anisotriang_susc_rescale}).

\section{\label{sec:conclusions}Conclusions}
In this paper we found the critical surface of the Ashkin-Teller model on the generic iso-radial graphs starting from 
the critical line of the anisotropic Ashkin-Teller model on the square lattice. Several special cases, such as 
the triangular and honeycomb lattice, were discussed in detail. Our method is in close connection with 
the duality relation and integrability of the critical surfaces. In other words, the  critical surfaces that we considered in our work
 on the square lattice
are self-dual and also integrable. Unfortunately they are not scanning the full critical surfaces in the anisotropic case: 
it seems, indeed, that there are regions of the full critical surface that are neither integrable nor self-dual even on the square lattice \cite{Tsallis}.
On the extracted critical surface for the Ashkin-Teller model on the generic iso-radial graph, we also found the free energy.

In the special case of the anisotropic triangular lattice, we checked our conjecture numerically and found an excellent 
agreement with the conjectured behaviour.
In addition we studied the lattice spinor variables for the Ising model 
in the non-critical case, finding the corresponding lattice equation of motions 
for Majorana fermions. It seems that a similar proceeding is possible for some particular 
points of the Ashkin-Teller model. Although we were not able to find the corresponding
 fermionic variables all along the critical line, we believe they exist. Another interesting 
study could be extracting the Thirring model, which is related to two Ising fermionic variables 
coupled by four-spinor interactions, with the same arguments; the difficulty in this
 direction might arise from writing down the multiplication of three disorder variables, which 
doesn't seem an immediate task.

From the numerical point of view there are different ways to expand our study. For example, 
one can check the validity of our conjecture in the portion of parameter space beyond the Ising model, especially 
in the triangular-lattice case where it seems, in contrast with the square lattice, that one can  
reach asymptotically the $X$-$Y$ point. Also, it could be interesting to explore the critical (anisotropic) Ising lines 
on different lattices; since they do not fall on the exactly solvable surfaces, there are no conjectures
 for their form, however it could be worth to see (at least numerically) their behaviour under a change of anisotropy angle.
Another interesting direction is an investigation of the fractal properties of critical interfaces in the isotropic Ashkin-Teller model \cite{fractal_interfaces, picco_santachiara_2010}.

\appendix
\section{\label{app:triang_honeycomb}Ashkin-Teller model on triangular and honeycomb lattices}
In this Appendix further details about special points on the (isotropic) triangular- and honeycomb-lattice critical lines are provided. A startling feature of the phase space is that the boundaries of the physical region, i.~e.~that with non-negative Boltzmann weights, do not respect universality, so that different geometries allow to span different ranges of universality classes within the physical region (Fig.~\ref{fig:at_critical_lines}). In the $x_1$-$x_2$ plane, this means that the physical threshold, $x_2=-1+2x_1$, crosses the three critical lines, Eqs.~\ref{eq:critline_x1x2_square}, \ref{eq:critline_x1x2_triangular} and \ref{eq:critline_x1x2_honeycomb}, at the already mentioned points $T^\Box$, $XY^\triangle$ and $U^{\hexagon}$ respectively, each one in its own universality class. In the $\beta,\alpha$ plane, this means that those are the rightmost points that can be expressed with real values of the couplings (by sending them to infinity while keeping the sum $\beta+\alpha$ fixed to a specific value). As already observed (Eq.~\ref{eq:XY_point_square_ba}), any attempt at translating non-physical points to this plane yields complex couplings.

The most advantageous case is the triangular lattice: here, the whole critical line falls in the physical region, the $XY$ point lying exactly on the boundaries; the coordinates for the relevant points are listed in Table \ref{tab:triangular_points}.
\begin{table}
	\begin{center}
	\begin{tabular}{|c||c|c|}
	\hline
	$\triangle$ \textit{point} & $\beta,\alpha$ coordinates & $x_1,x_2$ coordinates \\
	\hline
	\hline
	4-state Potts & $\beta=\alpha=\frac{\log 2}{4}$ & $x_1=x_2=\frac{1}{5}$ \\
	\hline
	F-Z & $\beta=-\frac{\log(2\sqrt{3}-3)}{4}\;,\; \alpha=\frac{\log 3}{8}$ & $x_1=\frac{(\sqrt{3}-1)(\sqrt{2}-1)}{\sqrt{2}}\;,\; x_2=3-2\sqrt{2}$ \\
	\hline
	Ising & $\alpha=0\;,\;\beta=\frac{\log 3}{4}$ & $x_1=2-\sqrt{3}\;,\;x_2=7-4\sqrt{3}$ \\
	\hline
	$XY$ & $\beta=\frac{\log 2}{2}-\alpha \to +\infty$ & $x_1=\sqrt{2}-1\;,\;x_2=2\sqrt{2}-3$ \\
	\hline
	\end{tabular}
	\caption{Relevant critical points for the triangular isotropic lattice.}
	\label{tab:triangular_points}
	\end{center}
\end{table}

The square lattice has already been presented: it falls within the physical region only up to the $T$ point, Eq.~\ref{Terminator_point_square_x}. The situation worsens for the honeycomb lattice: there, the critical line falls off the physical region starting from a point labelled $U$, that realises the following universality class: $\lambda_U^H=\frac{3}{5}\pi$, $\nu_U^H=\frac{4}{3}$ and $r_U^H=\sqrt{\frac{8}{5}}$. Table \ref{tab:honeycomb_points} contains the coordinates for the interesting points in this geometry.
\begin{table}
	\begin{center}
	\begin{tabular}{|c||c|c|}
	\hline
	$\hexagon$ \textit{point} & $\beta,\alpha$ coordinates & $x_1,x_2$ coordinates \\
	\hline
	\hline
	4-state Potts & $\beta=\alpha=\frac{\log 5}{4}$ & $x_1=x_2=\frac{1}{2}$ \\
	\hline
	F-Z & $\beta=-\frac{\log(3-2\sqrt{2})}{4}\;,\; \alpha=-\frac{\log(2-\sqrt{3})}{4}$ & $x_1=\frac{\sqrt{3}-1}{\sqrt{2}}\;,\; x_2=\sqrt{3}(2-\sqrt{3})$ \\
	\hline
	Ising & $\alpha=0\;,\;\beta=\frac{\log(2+\sqrt{3})}{2}$ & $x_1=\frac{1}{\sqrt{3}}\;,\;x_2=\frac{1}{3}$ \\
	\hline
	$U$ & $\beta=\frac{\log(1+\sqrt{5})}{2}-\alpha \to +\infty$ & $x_1=\frac{\sqrt{5}-1}{2}\;,\;x_2=\sqrt{5}-2$ \\
	\hline
	$XY$ & -- & $x_1=\frac{1}{2}\;,\;x_2=0$ \\
	\hline
	\end{tabular}
	\caption{Relevant critical points for the honeycomb isotropic lattice.}
	\label{tab:honeycomb_points}
	\end{center}
\end{table}

Duality provides correspondencies between points on the triangular and honeycomb lattices sharing the same universality class; on the other hand, the boundary of the physical region cuts the various critical lines in different points: this creates an interesting situation, in which a range of critical points (i.~e.~those with $\lambda>\lambda_U^H$) fall in the physical region on the triangular lattice but do not on the dual honeycomb realisation\footnote{More precisely, this happens for all points in the (triangular-lattice) phase space, including the ones off criticality, that lie below the curve $\alpha=-\frac{1}{2}\log\cosh 2\beta$ but still within the physical region.}. As already seen, the latter would correspond to values of $\beta,\alpha$ with a nonzero imaginary component $\pm\frac{\pi}{4}$. Nevertheless, duality must be respected even though, apparently, the Boltzmann weights are not always positive numbers. Indeed, a careful investigation reveals that the link-specific Boltzmann weights combine in such a way that, due to the geometric shape of the lattice and some constraints on the allowed spin configurations, the total weight for a given configuration is always a real number, and the resulting partition function is a well-defined, real and positive object (apart from a trivial overall prefactor).

This situation can also be considered from a somewhat reversed point of view: starting from a statistical model on the honeycomb lattice with complex couplings and therefore ill-defined Boltzmann weights, by making use of duality one can re-express all of the observables in a triangular-lattice version of the model, with well-defined, real and positive Boltzmann weights; moreover, the latter version of the original problem would now be open to numerical investigation. That would be the case, for instance, if one wanted to study the point with $r^2=\frac{3}{2},\;\lambda=\frac{5}{8}\pi,\;\nu=\frac{3}{2}$, i.~e.~the one corresponding to the twisted $\mathcal{N}=2$ supersymmetric CFT; the triangular lattice is, moreover, the only way to perform numerical investigation in the vicinity of the $XY$ point, which is accessible exclusively on this regularisation.

\section{\label{app:anisotropic_triangular}Anisotropic triangular lattice: complementary formulas}\label{app:anisotriang_gradients}
This Appendix provides some more formulas specific for the anisotropic triangular lattice with two equal sides (Section \ref{sec:montecarlo_triangular}). From the parametrisation for the critical manifold, Eqs.~\ref{eq:happytree_expl_1} and \ref{eq:happytree_expl_2}, two linearly independent vectors tangent to the surface in a given point $P$ can be evaluated as:
\begin{eqnarray}
	t_1 &=& \frac{\de}{\de\beta_1}\Big(\beta_1,\alpha_1,G(\beta_1,\alpha_1),H(\beta_1,\alpha_1)\Big)\Big|_{(\beta_1,\alpha_1)=P}\;; \\
	t_2 &=& \frac{\de}{\de\alpha_1}\Big(\beta_1,\alpha_1,G(\beta_1,\alpha_1),H(\beta_1,\alpha_1)\Big)\Big|_{(\beta_1,\alpha_1)=P}\;,
\end{eqnarray}
where the vectors are written as \mbox{$(\beta_1,\alpha_1,\beta_2,\alpha_2)$}. The segments used to probe the susceptibility are then constructed along two orthogonal solutions $w_1,w_2$ to the system:
\begin{equation}
	t_1 \cdot w = t_2 \cdot w = 0\;.
\end{equation}
The above tangent vectors are explicitly given by:
\begin{eqnarray}
	t_1 &=&\Big(
		1,0,\frac{2\tanh2\beta_1}{p-1}, \\
		&& \frac{\sinh2\alpha_1}{\sqrt{1+q^2}}\Big[ \frac{2\cosh2G}{(p-1)\cosh2\beta_1} - \frac{\sinh2G\cosh2\beta_1}{\sinh^22\beta_1} \Big]
	\Big)\;;\nonumber \\
	t_2 &=& \Bigg(
		0,1,\frac{r}{(1-p^2)d},
		\frac{
			\sinh2G\cosh2\alpha_1+\frac{
				r \sinh2\alpha_1 \cosh2G
			}{
				(1-p^2)d
			}
		}{
			\sqrt{1+q^2}\sinh2\beta_1
		}
	\Bigg)\;,
\end{eqnarray}
where the following shorthands have been employed:
\begin{eqnarray}
	d(\beta_1) &=& \cosh^22\beta_1 \;; \\
	p(\beta_1,\alpha_1) &=& \frac{
		1-\sinh^22\beta_1+2\sinh^22\alpha_1-2\sinh2\alpha_1\cosh2\alpha_1
	}{d(\beta_1)}\;; \\
	q(\beta_1,\alpha_1,G) &=& \frac{\sinh2G \sinh2\alpha_1}{\sinh2\beta_1}\;; \\
	r(\alpha_1) &=& 2(2\sinh2\alpha_1\cosh2\alpha_1-2\sinh^22\alpha_1-1) \;.
\end{eqnarray}

Another complementary observation to this particular case of anisotropy concerns the limits $\beta_1\to 0$ and $\beta_2\to 0$, taken by staying on a fixed universality class; the former yields $\beta_2\to\infty$, while the latter just gives $\beta_1 \to \beta_1^{S}$, that is, the corresponding square-lattice couplings. This can be easily understood by considering that the two limits implement the removing of two or one of the sides of the triangle, respectively. In other words, the first limit gives a system that is just a bundle of disconnected one-dimensional systems (never getting a nonzero magnetisation), and the second results in a square lattice with isotropic couplings.

We now provide three explicit ``iso-$\lambda$'' lines within the critical manifold; the four-state Potts model is given by:
\begin{equation}
	\alpha_i = \beta_i\;\;,\;\;
	\beta_2 = \frac{1}{2}\arctanh(1-2\tanh(2\beta_1))\;;
\end{equation}
the F-Z line is expressed as:
\begin{equation}
	\sinh 2\alpha_i = \frac{1}{\sqrt{2}}\sinh2\beta_i\;\;,\;\;
	\beta_2 = \frac{1}{2} \arctanh\Big( \frac{1-\sinh2\beta_1\sqrt{2+\sinh^22\beta_1}}{\cosh^22\beta_1} \Big)\;;
\end{equation}
and finally the Ising sub-manifold is:
\begin{equation}
	\alpha_i = 0\;\;,\;\;
	\beta_2 = \frac{1}{2}\arctanh\Big( \frac{1-\sinh^22\beta_1}{\cosh^22\alpha_1} \Big)\;.
\end{equation}
These are found by imposing $\lambda = 0,\frac{\pi}{4},\frac{\pi}{2}$ in Eqs.~\ref{eq:happytree_expl_1} and \ref{eq:happytree_expl_2} by means of Eq.~\ref{critical_line_anisotropic_AT_on_the_anisotropic_triangular_3}.

\vspace{0.4cm}
\textit{Acknowledgments}:
We are grateful to M.~Caselle for helpful suggestions during the early stages of this work. We are also indebted to F.~Gliozzi for fruitful discussions.


\begin{thebibliography}{99}
\bibitem{Ashkin} J.~Ashkin and E.~Teller, Phys.~Rev.~\textbf{64}, (1943) 178.
\bibitem{DR} E.~Domany, E.~K.~Riedel, Phys.~Rev.~Lett.~\textbf{40}, (1978), 561-564.
\bibitem{bak} P.~Bak, P.~Kleban, W.~N.~Unertl, J.~Ochab, G.~Akinci, N.~C.~Bartelt and T.~L.~Einstein, Phys.~Rev.~Lett.~\textbf{54} (1985), 1539-1542.
\bibitem{BEW} N.~C.~Bartelt, T.~L.~Einstein, L.~T.~Wille, Phys.~Rev.~\textbf{B} 40 (1989), 10759.
\bibitem{wu_and_lin} F.~Y.~Wu and K.~Y.~Lin, J.~Phys.~\textbf{C}: Solid St.~Phys.~\textbf{7} (1974), L181; J.~Math.~Phys.~\textbf{17} (1976) 439.
\bibitem{Kadanoff_and_Brown} L.~P.~Kadanoff, A.~Brown, Ann.~Phys., NY, \textbf{121}, 318 (1979).
\bibitem{baxter} R.~J.~Baxter, \textit{Exactly Solved Models in Statistical Mechanics}, Associated Press, London, 1982, p.~471.
\bibitem{nienhuis2} B.~Nienhuis, J.~Stat.~Phys.~\textbf{34} (1984), 731-761.
\bibitem{Kohmoto_den_nijs} M.~Kohmoto, M.~den Nijs and L.~Kadanoff, Phys.~Rev.~\textbf{B} 24 (1981), 5229-5241.
\bibitem{DG} G.~Delfino, P.~Grinza, Nucl.~Phys.~\textbf{B} 682 (2004), 521-550 [arXiv:hep-th/0309129].
\bibitem{Den_nijs} M.~P.~M.~Den Nijs, Phys.~Rev.~\textbf{B} 23 (1981), 6111.
\bibitem{yang} S-K.~Yang, Nucl.~Phys.~\textbf{B} 285 [FS19] (1987), 183-203.
\bibitem{Ginsparg} P.~Ginsparg, Nucl.~Phys.~\textbf{B} 295 [FS21] (1988) 153-170; P.~Ginsparg, arXiv:hep-th/9108028.
\bibitem{Saleur} H.~Saleur, J.~Phys.~\textbf{A}: Math.~Gen.~20 (1987), L1127-L1133.
\bibitem{Schellekens} A.~N.~Schellekens, Fortschr.~Phys.~\textbf{44} (1996) 605.
\bibitem{DVVV} R.~Dijkgraaf, C.~Vafa, E.~Verlinde and H.~Verlinde,
Commun.~Math.~Phys.~\textbf{123} (1989) 485.
\bibitem{von_gehlen_and_rittenberg} \textit{For the connection of this model to superconformal field theory see}
G.~Von Gehlen and V.~Rittenberg,
J.~Phys.~\textbf{A}: Math.~Gen.~20 (1987) 227-237; M.~Baake, G.~Von Gehlen and V.~Rittenberg, J.~Phys.~\textbf{A}:
Math.~Gen.~20 (1987) L479-L485.
\bibitem{zamolodchikov} A.~Zamolodchikov, Nucl.~Phys.~\textbf{B}, 285 (1987) 481-503.
\bibitem{schramm} O.~Schramm, Israel J.~Math.~\textbf{118} (2000), 221 [arXiv:math.PR/9904022].
\bibitem{Kager} W.~Kager, B.~Nienhuis, J.~Stat.~Phys.~\textbf{115} (2004), 1149-1229 [arXiv:math-ph/0312056];
J.~Cardy, Ann.~Phys.~\textbf{318} (2005), 81-118 [arXiv:cond-mat/0503313];
M.~Bauer, D.~Bernard, Phys.~Rept.~\textbf{432} (2006), 115-221 [arXiv:math-ph/0602049].
\bibitem{CS} D.~Chelkak, S.~Smirnov, arXiv:0910.2045 (math-ph).
\bibitem{baxter2} R.~G.~Baxter, Phil.~Trans.~R.~Soc.~Lond.~\textbf{A} 289 (1978), 315.
\bibitem{baxter3} R.~G.~Baxter, Proc.~R.~Soc.~Lond.~\textbf{A} 404 (1986), 1.
\bibitem{mercat} C.~Mercat, Comm.~Math.~Phys.~\textbf{218} (2001), 177.
\bibitem{ZF} A.~B.~Zamolodchikov, V.~A.~Fateev, Sov.~Phys.~JETP \textbf{62} (1985), 215;
Phys.~Rev.~\textbf{B}, 24, 9 (1981), 5229-5241.
\bibitem{RC} M.~A.~Rajabpour, J.~Cardy, J.~Phys. \textbf{A}: Math.~Theor.~40 (2008), 14703-14713 [arXiv:0708.3772 (math-ph)].
\bibitem{BPP} M.~N.~Barber, I.~Peschel and P.~A.~Pearce, J.~Stat.~Phys.~\textbf{37} (1984), 497; D.~Kim and P.~Pearce, J.~Phys.~\textbf{A} 20 (1987), L451.
\bibitem{Dot} V.~S.~Dotsenko, V.~S.~Dotsenko, Advances in Physics \textbf{32} (1983), 129-172.
\bibitem{KS}  R.~Kenyon and J.-M.~Schlenker, Trans.~Amer.~Math.~Soc.~\textbf{357} (2005), 3443.
\bibitem{SP} P.~A.~Pearce, J.~Phys.~\textbf{A}: Math.~Gen.~20 (1987), 6463-6469; K.~A.~Seaton, P.~A.~Pearce, J.~Phys.~\textbf{A}: Math.~Gen.~22 (1989), 2567-2576.
\bibitem{TA} H.~N.~V.~Temperley, S.~E.~Ashley, Proc.~R.~Soc.~Lond.~\textbf{A} 365 (1979), 371-380.
\bibitem{BMS} A.~I. Bobenko, C.~Mercat, Y.~B. Suris, J.~Reine Angew.~Math.~\textbf{583} (2005), 117-161 [arXiv:math/0402097].
\bibitem{Zamol} A.~B.~Zamolodchikov, Communications in Mathematical Physics \textbf{69} (1979), 165-178.
\bibitem{BP} S.~V.~Pokrovsky and Yu.~A.~Bashilov, Communications in Mathematical Physics \textbf{84} (1982), 103-132.
\bibitem{z4_gauge_dualAT} P.~Giudice, F.~Gliozzi, S.~Lottini, JHEP \textbf{01} (2007) 084 [arXiv:hep-th/0612131].
\bibitem{z4_gauge} M.~Caselle, P.~Giudice, F.~Gliozzi, P.~Grinza, S.~Lottini, JHEP \textbf{11} (2007) 075 [arXiv:0710.3522 (hep-lat)].
\bibitem{Tsallis} C.~Tsallis and J.~Souletie, J.~Phys.~\textbf{A}: Math.~Gen.~19 (1986) 1715-1725.
\bibitem{fractal_interfaces} M.~Caselle, S.~Lottini, M.~A.~Rajabpour, arXiv:0907.5094 (cond-mat.stat-mech).
\bibitem{picco_santachiara_2010} M.~Picco, R.~Santachiara, \textit{in preparation}.
\end{thebibliography}
\end{document}